\documentclass{article}
\usepackage{graphicx} % Required for inserting images

\usepackage{PRIMEarxiv}

\usepackage[utf8]{inputenc} % allow utf-8 input
\usepackage[T1]{fontenc}    % use 8-bit T1 fonts
\usepackage{hyperref}       % hyperlinks
\usepackage{url}            % simple URL typesetting
\usepackage{booktabs}       % professional-quality tables
\usepackage{amsfonts}       % blackboard math symbols
\usepackage{nicefrac}       % compact symbols for 1/2, etc.
\usepackage{microtype}      % microtypography
\usepackage{lipsum}
\usepackage{fancyhdr}       % header
\usepackage{graphicx}       % graphics
\graphicspath{{media/}}     % organize your images and other figures under media/ folder
\usepackage{amsmath}
\usepackage{algorithm}
\usepackage{algorithmicx}
\usepackage{algpseudocode}
\usepackage{multirow}
\usepackage{todonotes}
\usepackage{comment}
\usepackage{caption}
\usepackage{subcaption}
\usepackage{amsthm}

\newtheorem{lemma}{Lemma}

\newtheorem{definition}{Definition}

\newcounter{subdefinition}[definition]
\renewcommand{\thesubdefinition}{\thedefinition.\arabic{subdefinition}}

\newcommand{\paperTitle}{\textsf{Sarus}: Privacy-Preserving Multi-Vendor Perception Fusion via Homomorphic Encryption}

\title{\paperTitle}
\author{
Munawar Hasan$^{1,2}$,
Apostol Vassilev$^{1}$ \\
\\
$^{1}$National Institute of Standards and Technology, USA \\
$^{2}$Michigan Technological University, USA \\
\\
\texttt{\{munawar.hasan, apostol.vassilev\}@nist.gov} \\
\texttt{munawarh@mtu.edu}
}
\begin{document}

\maketitle

\begin{abstract}
Cooperative perception enables autonomous vehicles (AVs) to improve situational awareness by aggregating detection outputs from multiple agents and sensing platforms, often via a shared fusion service in multi-vendor deployments. However, sharing such outputs at inference time exposes proprietary model behavior and sensitive environmental information, creating significant privacy and security concerns. In this paper, we present \textsf{Sarus}, a privacy-preserving framework for multi-vendor perception fusion via homomorphic encryption (HE), enabling aggregation without revealing individual vendor outputs. Each vendor encodes detections as compact Gaussian moment vectors over a shared spatial lattice and transmits encrypted payloads to a fusion server, which aggregates them directly in the encrypted domain. The fused result is then decrypted and reconstructed into final detections through class-wise bin merging.

We analyze the computational complexity, showing linear scaling for vendor payload construction and $O(BV)$ server-side fusion with the number of occupied bins $B$ and vendors $V$, while postprocessing scales as $O(B + \sum_{c\in \mathcal{C}} B_c^2)$, where $\mathcal{C}$ denotes the set of object classes and $B_c$ is the number of occupied bins for class $c$. Experiments demonstrate linear scaling in practice with only a bounded constant-factor overhead from HE, with decryption dominating postprocessing cost. Experiments on the KITTI dataset using camera (YOLOv8 \footnote{Certain equipment, instruments, software, or materials, commercial or noncommercial, are identified in this paper to specify the experimental procedure adequately. Such identification does not imply recommendation or endorsement of any product or service by NIST, nor does it imply that the materials, equipment, or software identified are necessarily the best available for the purpose.}) and LiDAR (PointPillars, PV-RCNN) detectors show that \textsf{Sarus} improves scene-level coverage by effectively aggregating complementary detections, particularly in distance-dependent regimes where individual modalities degrade. These results indicate that privacy-preserving multi-vendor perception fusion is feasible for real-time deployment when statistical compression and spatial sparsity are jointly exploited. The demonstration code and dataset for this project is open source, available on Github\footnote{\url{https://github.com/mhasan08/sarus}}.
\end{abstract}


\section{Introduction}
\label{sec:introduction}
Cooperative perception enables multiple agents, such as autonomous vehicles, roadside infrastructure, and perception service providers, to share complementary observations of a common environment. By aggregating information from diverse viewpoints and sensing modalities, cooperative perception improves robustness under occlusion, extends sensing range beyond line-of-sight, and enhances reliability in challenging conditions such as adverse weather or dense traffic~\cite{chen2019cooperatively, wang2020v2vnet, xu2021opencood, xiang2023multi, kim2014multivehicle}. Beyond spatial diversity, cooperative perception also leverages \textit{temporal and contextual diversity}, where observations collected across time or from different agents contribute to a more stable and consistent understanding of dynamic scenes. This enables improved tracking of objects, early detection of hazards, and more reliable estimation of scene structure compared to isolated perception systems. Cooperative perception transforms perception from a purely local task into a \textit{distributed sensing and aggregation problem}, where multiple heterogeneous sources jointly contribute to a unified representation of the environment. 

In practical deployments, cooperative perception is often realized through a centralized or edge-based \textit{fusion server}~\cite{dai2020hybrid, boehme2020talkycars}, which collects perception outputs from multiple sources and aggregates them into a unified scene representation. Each participant processes its local sensor data independently and transmits a compact representation—such as detected objects, bounding boxes, or intermediate features—to the fusion server. The server then combines these inputs to produce a consistent and enriched understanding of the environment, which can be redistributed to participating agents or used for downstream decision-making. Multi-modal fusion further reinforces the role of the fusion server as an aggregation point for heterogeneous data representations. Each vendor or agent may process distinct sensor modalities or employ modality-specific models, producing outputs that differ in structure, scale, and uncertainty characteristics. The fusion process must therefore reconcile these heterogeneous representations into a consistent spatial interpretation of the scene. By integrating these distributed observations, the fusion server enables a more complete and accurate representation of the scene than any single agent could achieve independently. Real-world deployments, such as the University of Michigan Smart Intersection Project, demonstrate this paradigm by aggregating multi-sensor observations to construct real-time intersection-level scene understanding~\cite{umtri_sip}. This paradigm shifts perception from isolated sensing to distributed scene understanding.

\begin{figure}[!t]
\centering

\begin{subfigure}[t]{0.48\textwidth}
    \centering
    \begin{minipage}[c][5.1cm][c]{\linewidth}
        \centering
        \includegraphics[width=\linewidth]{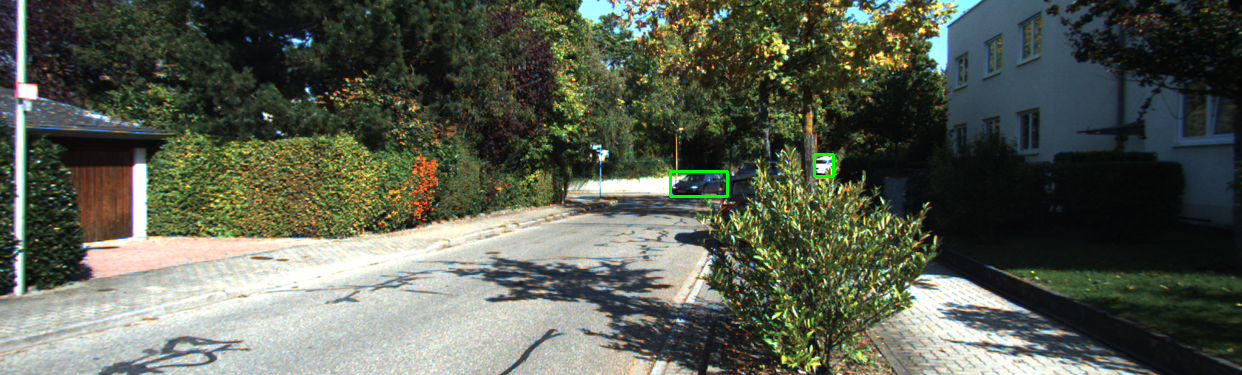}
    \end{minipage}
    \caption{Camera-only perception: YOLOv8 detects two cars.}
    \label{fig:yolov8}
\end{subfigure}
\hfill
\begin{subfigure}[t]{0.51\textwidth}
    \centering
    \begin{minipage}[c][5.1cm][c]{\linewidth}
        \centering
        \includegraphics[width=\linewidth]{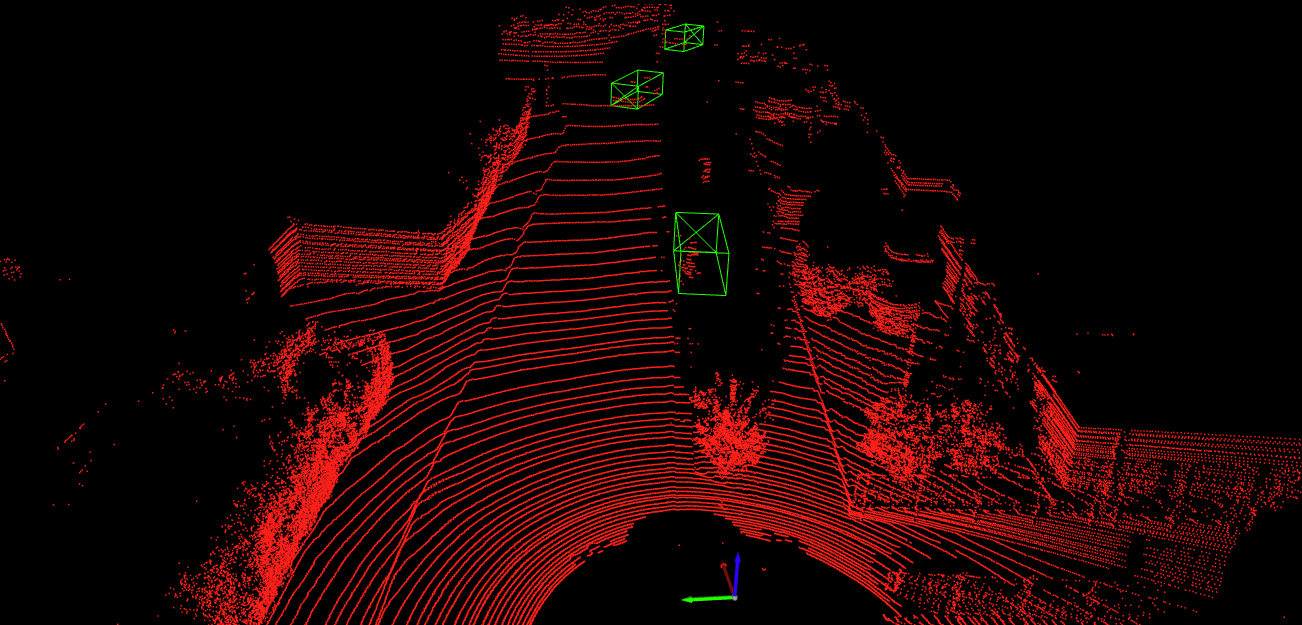}
    \end{minipage}
    \caption{LiDAR-only perception: PV-RCNN. detects three cars.}
    \label{fig:pvrcnn}
\end{subfigure}

\caption{Complementarity of heterogeneous perception outputs on KITTI (Karlsruhe Institute of Technology and Toyota Technological Institute)~\cite{kitti}: The camera detector captures visually salient vehicles, while the LiDAR detector recovers additional 3D structure, including a nearby vehicle missed in the image due to partial occlusion. This example motivates privacy-preserving cooperative fusion: multiple agents or vendors may collectively improve scene coverage, but direct sharing of raw sensor data or detailed proprietary outputs can expose sensitive information.}
\label{fig:kitti_detection}
\end{figure}

Importantly, cooperative perception operates at the \textit{inference level}, where agents exchange the outputs of their local perception pipelines rather than raw sensor data or model parameters. Each participant independently processes its sensor inputs and produces detection results—such as object locations, classes, and confidence scores—which are then shared for aggregation. This design avoids the need for centralized training or access to raw data, making it well-suited for real-time deployment in dynamic environments. In Figure~\ref{fig:kitti_detection}, YOLOv8~\cite{redmon2016you} detects two vehicles in the camera image, while PV-RCNN~\cite{pvrcnn} identifies three using LiDAR (Light Detection and Ranging) data. The missed detection by YOLOv8 corresponds to the closest vehicle, which is occluded by vegetation. This example highlights how inference-level cooperative perception can leverage complementary modalities to recover missed objects. However, operating at the inference level introduces unique challenges. Unlike training-time collaboration frameworks such as federated learning, where model updates can be aggregated without exposing individual predictions, cooperative perception requires sharing fine-grained outputs that directly reflect model behavior on specific scenes. These outputs inherently encode both environmental information and model-specific characteristics, making them highly sensitive. As a result, protecting inference-time data becomes critical for preserving privacy, confidentiality, and competitive integrity in multi-vendor settings.

Despite its advantages, cooperative perception introduces a fundamental challenge between \textit{utility and privacy}. Vendor detections encode not only information about the environment, but also implicit knowledge about proprietary models, training data, and system behavior. Sharing raw perception outputs in cooperative perception systems exposes multiple security and privacy risks. In particular, plaintext detection sharing enables the following attack vectors:
\begin{enumerate}
    \item \textbf{Model Extraction:} Repeated access to detection outputs allows an adversarial aggregation service or participating vendor to infer properties of proprietary perception models. By observing confidence scores, bounding box behavior, and response patterns across diverse inputs, an attacker can approximate model decision boundaries or reconstruct surrogate models~\cite{vassilev2025, chen2020improved, wu2016methodology}. This compromises vendor intellectual property. Further, such compromise can also reveal vendor-specific design choices and performance characteristics, enabling competitors to reverse-engineer or benchmark proprietary systems.
    \item \textbf{Data Leakage at Inference:} Detection outputs may reveal sensitive information about the observed environment, including the presence, location, and behavior of objects or individuals. In infrastructure-assisted or multi-agent settings, sharing such outputs in plaintext can inadvertently expose private or regulated data~\cite{vassilev2025, dibbo2023sok}.
    \item \textbf{Side-Channel Leakage:} Auxiliary information such as timing, output sparsity, and detection patterns can leak additional information beyond the explicit content of detections. For example, variations in detection latency or confidence distributions may reveal scene complexity, sensor characteristics, internal model behavior, or model format~\cite{vassilev2025, wiz} further amplifying privacy risks.
    \item \textbf{Cross-Vendor Inference.} In multi-vendor settings, access to multiple vendors' outputs for the same scene enables comparative analysis, allowing adversaries to infer relative strengths, weaknesses, or biases of individual models. Such cross-analysis can expose competitive information and facilitate targeted attacks against specific vendors.
\end{enumerate}

This challenge is further amplified in \textit{multi-vendor} settings, where participating entities may be competitors or operate under distinct trust and regulatory constraints. Existing cooperative perception systems largely assume access to raw or intermediate detection outputs in plaintext, making them incompatible with privacy-preserving deployment scenarios. Consequently, there is a pressing need for mechanisms that enable \textit{secure aggregation of perception outputs} without revealing vendor-specific information.

In this paper, we present \textsf{Sarus}\footnote{\textsf{Sarus} draws inspiration from the Sarus crane—a species native to India and Southeast Asia—known for its coordinated movement and broad environmental awareness, mirroring the collaborative perception paradigm underlying our framework.}, a privacy-preserving framework for multi-vendor perception fusion based on homomorphic encryption (HE). Instead of sharing raw detections, each vendor encodes its observations into compact \textit{moment-based statistical representations}. In this representation, detections are summarized through aggregate quantities such as confidence mass, confidence-weighted object center, and spatial spread, which capture object location, uncertainty, and confidence without exposing the raw detection outputs. These representations are encrypted using CKKS (Cheon--Kim--Kim--Song)~\cite{cheon2017homomorphic} and transmitted to a fusion server, which performs aggregation directly in the encrypted domain without accessing plaintext data. The fused result is then decrypted by an authorized party and used to reconstruct the final detections.

A key aspect of our design is the use of a shared spatial lattice that discretizes the scene into bins, enabling scalable aggregation without explicit cross-vendor matching. By combining moment-based encoding with spatial binning, \textsf{Sarus} achieves linear scaling in the number of occupied bins while maintaining robustness to minor spatial misalignment. Importantly, the use of homomorphic encryption introduces only a bounded constant-factor overhead, preserving practical feasibility. We summarize our contributions below:
\begin{itemize}
    \item We introduce a homomorphically encrypted moment-based fusion framework that enables aggregation of multi-vendor perception outputs without revealing raw detections.
    \item We propose a spatial binning mechanism that ensures scalable fusion with complexity $O(B \cdot V)$, where $B$ is the number of occupied bins and $V$ the number of vendors.
    \item We provide a comprehensive performance evaluation covering vendor payload generation, server-side homomorphic fusion, and postprocessing.
    \item We demonstrate that homomorphic encryption incurs only bounded overhead, with linear scaling in practice and decryption dominating vendor-side cost.
    \item We show empirically on KITTI~\cite{kitti} that \textsf{Sarus} effectively recovers complementary detections across camera and LiDAR modalities. 
\end{itemize}

Our results show that privacy-preserving multi-vendor perception fusion is feasible in real-time settings when statistical compression and spatial sparsity are jointly exploited, enabling secure and scalable cooperative perception in heterogeneous environments. We clarify that \textsf{Sarus} does not handle authentication, identity management, schema validation, or proof-carrying compliance for cooperative perception. Instead, it assumes that submitted payloads have passed an admission check before encrypted fusion. This admission check includes both agent legitimacy and adherence to a shared syntax and public specification for the submitted perception output. Such an admission layer may be built from proof-carrying data~\cite{chiesa2010pcd}, vehicular message security and credential mechanisms such as IEEE~1609.2~\cite{ieee1609}, and succinct zero-knowledge compliance mechanisms such as Hermes Seal~\cite{hasan2026hermessealzeroknowledgeassurance}.

\section{Related Work}
\label{sec:related-work}

Cooperative perception enables multiple agents to share observations and improve scene understanding beyond the capabilities of individual sensors. Prior work has explored vehicle-to-vehicle (V2V) and vehicle-to-infrastructure (V2I) collaboration for object detection and tracking, demonstrating improved robustness under occlusion and extended sensing range~\cite{chen2019cooperatively, wang2020v2vnet, xu2021opencood}. These approaches typically rely on exchanging raw sensor data~\cite{xiang2023multi, zheng2022robust, arnold2020cooperative}, intermediate features~\cite{xu2022v2x, ma2024macp, rauch2012car2x}, or detection outputs~\cite{li2023learning}, and assume a trusted setting in which such information can be shared in plaintext. Infrastructure-assisted systems further extend this paradigm by aggregating observations from roadside units and vehicles to construct a unified scene representation, often via a centralized fusion module. Modern perception systems increasingly incorporate multiple sensing modalities, such as cameras, LiDAR, and radar, to improve robustness and accuracy~\cite{wei2025cooperative}. Multi-modal fusion methods~\cite{zhang2024collaborative, zhou2024vit, he2025mdnet, yin2023v2vformer++, zhang2022multi, li2025rg, lu2025privacy} combine complementary information from heterogeneous sensors to enhance detection performance under challenging conditions~\cite{xiang2023multi, jiang2025multimodal}. In cooperative settings, this heterogeneity is further amplified, as different agents may employ distinct sensor configurations and model architectures. Existing fusion strategies typically operate on feature-level or detection-level representations and require direct access to underlying data, making them difficult to deploy in privacy-sensitive multi-vendor environments.

Privacy-preserving machine learning techniques, such as federated learning (FL)~\cite{mcmahan2017communication, kairouz2021advances, karimireddy2020scaffold, mammen2021federated}, enable collaborative model training without sharing raw data. In FL, participants exchange model updates rather than inference outputs, reducing exposure of local datasets. However, FL may not be suitable for cooperative perception since, FL cannot operate at \textit{inference time}, requiring aggregation of instance-level predictions rather than model parameters~\cite{zhang2024federated, abdel2021vehicular}. As a result, existing FL-based approaches do not directly address the privacy risks associated with sharing detection outputs across vendors. Secure aggregation protocols and homomorphic encryption  have been widely studied for enabling computation over encrypted data~\cite{hesamifard2017cryptodl, xu2019cryptonn, cryptoeprint:2014/331}. HE schemes, such as CKKS~\cite{cheon2017homomorphic}, support approximate arithmetic on encrypted vectors and have been applied in domains such as secure inference and privacy-preserving analytics~\cite{moon2025thor, kim2023optimized}. Prior work has explored encrypted aggregation in distributed learning and statistics; however, these approaches typically focus on scalar or low-dimensional data and do not address the challenges of structured perception outputs, such as spatial alignment, object-level aggregation, and multi-modal heterogeneity.

Existing cooperative perception systems typically assume plaintext sharing of detection outputs, while privacy-preserving learning approaches focus on training or intermediate representations rather than inference-time collaboration. While prior work has explored cooperative perception and privacy-preserving computation independently, a scalable framework for privacy-preserving multi-vendor perception fusion at inference time remains largely unexplored, particularly for structured detection outputs. \textsf{Sarus} bridges this gap by introducing an encrypted moment-based representation and a spatial binning strategy that enable scalable homomorphic aggregation of structured perception outputs without revealing vendor-specific information.

\section{Data Representation, System, and Threat Model}
\label{sec:preliminaries}
We start by presenting notations used in this paper. Table~\ref{tab:abbreviations} summarizes the abbreviations and mathematical notations.
\begin{table}[H]
    \centering
    \caption{Symbols and Mathematical Notations}
    \label{tab:abbreviations}
    \begin{tabular}{|l|p{10cm}|}
        \hline
        $\mathcal{V}$ & Vendors, $\mathcal{V} \in \{v_1 \dots v_V\}$ such that $|\mathcal{V}| = V$\\
        \hline
        $\mathsf{Enc}_{\mathsf{HE}}(\cdot)$,  $\mathsf{Dec}_{\mathsf{HE}}(\cdot)$ & Homomorphic encryption and decryption\\
        \hline
        $\mathcal{P}_v$ & Encrypted payload from vendor $v \in \mathcal{V}$\\
        \hline
        $\mathcal{C}$ & Set of object classes\\
        \hline
        $\mathcal{F}_{\text{common}}$ & Common spatial reference frame.\\
        \hline
        $\mathbf{b}$ & Bounding box with coordinates $x_1, y_1, x_2, y_2$.\\
        \hline
        $(w, h)$ & Width and height of bounding box.\\
        \hline
        $\mu$, $\sigma^2$ & Mean and variance.\\
        \hline
        $\alpha$ & Vendor trust weight.\\
        \hline
    \end{tabular}
\end{table}

\subsection{Data Representation}
\begin{definition}[Common Spatial Frame]
\label{def:common_frame}
Let $\mathcal{F}_{\text{common}} \subseteq \mathbb{R}^d$ denote a shared spatial reference frame in which detections from all vendors are expressed. Let $\mathcal{O}$ denote a physical object in the environment, and for each vendor $v \in \mathcal{V}$, let $\Phi_v$ denote the observation operator that maps $\mathcal{O}$ to the vendor's sensor observation $\mathcal{I}_v$, i.e., $$\Phi_v : \mathcal{O} \rightarrow \mathcal{I}_v.$$
Further, let $\mathcal{T}_v$ denote a transformation that maps observations to the common spatial frame, then we have:
$$\mathcal{T}_v : \mathcal{I}_v \rightarrow \mathcal{F}_{\text{common}}.$$
We assume that for any object $\mathcal{O}$ observed by multiple vendors, the composed mappings $\mathcal{T}_v \circ \Phi_v(\mathcal{O})$ are spatially consistent, i.e., they lie within a bounded neighborhood in $\mathcal{F}_{\text{common}}$.
\end{definition}
$\Phi_v$ captures sensing and perception effects (e.g., viewpoint, modality, and model behavior), while $\mathcal{T}_v$ accounts for geometric alignment through calibration and localization.

\begin{definition}[Spatial Consistency Assumption]
\label{def:spatial_consistency}
For any physical object $\mathcal{O}$ observed by vendors $v$ and $v'$, the corresponding coordinates $(x_v, y_v)$ and $(x_{v'}, y_{v'})$ in $\mathcal{F}_{\text{common}}$ satisfy: $$
\|(x_v, y_v) - (x_{v'}, y_{v'})\| \leq \epsilon,$$ for some bounded alignment error $\epsilon > 0$ determined by calibration and localization accuracy.
\end{definition}

The common spatial frame $\mathcal{F}_{\text{common}}$ can take several forms depending on the deployment setting. In many autonomous driving systems, detections are projected into a bird’s-eye-view (BEV) representation aligned with the ground plane, enabling consistent spatial reasoning across agents. Alternatively, an ego-centric coordinate frame centered at a reference vehicle may be used, where all detections are expressed relative to the vehicle’s pose. In infrastructure-assisted settings, detections may be aligned to a global or map-based coordinate system using GPS and localization pipelines. These transformations are standard in cooperative perception systems and rely on well-established calibration and pose estimation techniques.

\subsection{System Model}
\label{sec:system_model}

We consider a multi-vendor cooperative perception setting consisting of a set of vendors $\mathcal{}$ such that $|\mathcal{V}| = V$ and a centralized fusion server. Each vendor $v \in \mathcal{V}$ operates an independent perception pipeline over its local sensor observations and participates in collaborative inference through secure aggregation.

\paragraph{Vendors:}
Each vendor $v$ observes the environment through local sensors, producing observations $\mathcal{I}_v$ (e.g., images or point clouds). Using its proprietary perception model, the vendor extracts detection outputs and encodes them into a compact moment-based representation over a shared spatial frame $\mathcal{F}_{\text{common}}$. The resulting payload is encrypted using a homomorphic encryption scheme $\mathsf{Enc}_{\mathsf{HE}}(\cdot)$ before transmission.

\paragraph{Fusion Server:}
The fusion server receives encrypted payloads from the participating vendors and performs aggregation directly in the encrypted domain. Its role is limited to computing aggregated statistics over ciphertexts; it does not require access to plaintext detections, model parameters, or intermediate vendor representations.

\paragraph{Output Reconstruction:}
After aggregation, the encrypted fused result is returned to an authorized party (e.g., a participating vendor or a trusted client), which decrypts the payload using $\mathsf{Dec}_{\mathsf{HE}}(\cdot)$ and reconstructs final detections through spatial and statistical postprocessing.

\paragraph{Communication Model:}
Communication occurs in two stages: (i) vendors transmit encrypted payloads to the fusion server, and (ii) the fusion server returns an aggregated encrypted result. Note, no plaintext perception outputs are exchanged between vendors or revealed to the server.

\paragraph{Assumptions:}
We assume that all vendors express detections in a shared spatial reference frame $\mathcal{F}_{\text{common}}$ (see Definition~\ref{def:common_frame}). Vendors do not share raw observations or model parameters, and encryption keys are not accessible to the server.

\subsection{Threat Model}
\label{sec:threat_model}

Building on the system model in Section~\ref{sec:system_model},  Our threat model focuses on protecting the confidentiality of vendor perception outputs at inference time.

\paragraph{Adversarial Fusion Server:}
\textit{Honest-but-curious}---It correctly follows the prescribed aggregation protocol but may attempt to infer information about individual vendor inputs from the received data and intermediate computations. In particular, the server has access to all transmitted payloads and aggregation results, and may perform arbitrary offline analysis on observed data. The server does not possess decryption keys and cannot directly access plaintext representations.

\paragraph{Curious Vendors:}
\textit{Mutually untrusted}--- May attempt to infer information about other participants through the fusion process. However, vendors only observe their own inputs and the final fused output after decryption, and do not have access to intermediate per-vendor contributions.

\paragraph{Adversarial Capabilities:}
The adversary may --- (i) observe all encrypted payloads transmitted to the fusion server, (ii) access aggregated ciphertexts and final fused outputs, and (iii) perform statistical or inference attacks on observed data. The adversary is not able to--- (i) break the underlying homomorphic encryption scheme, (ii) access secret keys or decrypt intermediate ciphertexts, and (iii) tamper with the protocol execution (i.e., no active attacks).

\paragraph{Security Goals:}
Our objective is to ensure that no party learns any information about individual vendor detections beyond what is revealed by the final fused output. In particular, we aim to:
\begin{itemize}
    \item Protect the confidentiality of vendor detection outputs, including object locations and confidence scores.
    \item Prevent inference of proprietary model behavior from shared data.
    \item Ensure that intermediate aggregation steps do not leak additional information.
\end{itemize}

\paragraph{Out of Scope:}
We assume that all admitted vendors express detections in a shared spatial reference frame $\mathcal{F}_{\text{common}}$ (see Definition~\ref{def:common_frame}) and follow the agreed payload schema and public fusion specification. Admission control, authentication, and compliance checking are treated as prerequisite mechanisms rather than functions provided by \textsf{Sarus}. 

We do not consider active adversaries that deviate from the protocol, collusion between multiple parties, or side-channel attacks arising from implementation-specific leakage (e.g., timing or hardware effects). These extensions are left for future work. A formal treatment of inference-time privacy with explicit leakage bounds can be developed under standard semantic security assumptions of homomorphic encryption scheme; we defer a detailed treatment to future work.

\paragraph{Intuition:}
At a high level, \textsf{Sarus} ensures that the fusion server operates exclusively on encrypted representations of vendor data. Since all vendor payloads are encrypted under a semantically secure homomorphic encryption scheme, the server cannot access individual detections or intermediate statistics in plaintext. Note that we use the term semantic security in its standard cryptographic sense: ciphertexts should not reveal any efficiently computable information about the underlying plaintext, except what is already implied by public information~\cite{goldwasser1984probabilistic,katz2020introduction}. Aggregation is performed through homomorphic operations, and only the final fused result is revealed after decryption. Thus, the server's view is computationally indistinguishable from that of an observer with access only to encrypted data and the final output.

\section{\textsf{Sarus}}
\label{sec:sarus}
The key challenge in privacy-preserving perception fusion is to aggregate object detections from multiple vendors without revealing proprietary model outputs or intermediate representations of those outputs. Sharing bounding boxes or prediction confidences of the perception stack in plaintext to vendors exposes sensitive perception information, therefore violating the trust-neutral collaboration model. To address this challenge, \textsf{Sarus} converts each detection into a representation whose sufficient statistics can be aggregated using only linear operations. This design enables the fusion server to combine perception outputs directly in encrypted form using homomorphic encryption, while preserving the geometric information required to reconstruct the fused detections.

Specifically, each bounding box is represented as a Gaussian splat and encodes its weighted spatial statistics as a moment vector. In this work, Gaussian splatting refers to representing each detection as a spatial Gaussian contribution over the common fusion grid, where the Gaussian center corresponds to the detected object location and the covariance/spread captures localization uncertainty. This allows detections from multiple vendors to be accumulated as smooth statistical evidence rather than as discrete bounding boxes. The moment vectors are accumulated within spatial bins and homomorphically aggregated across vendors. After decryption, the fused moments are inverted to recover the parameters of the fused Gaussian representation, which is then converted back into a bounding box estimate. This representation allows collaborative perception fusion while maintaining strict privacy guarantees for vendor detections.

\subsection{Detection Representation via Gaussian Splats}
\label{sec:detection-to-gaussian-splat}
Bounding boxes represent discrete geometric primitives whose fusion typically requires non-linear operations such as intersection tests, non-maximum suppression, or heuristic box merging strategies. Such operations are incompatible with secure aggregation under homomorphic encryption, which efficiently supports only linear computations. To address this limitation, \textsf{Sarus} converts each bounding box detection into
a continuous spatial representation using a \textit{Gaussian splat}.
The Gaussian representation models the detected object by its spatial center and an associated uncertainty that reflects the spatial extent of the bounding box. This representation provides two important advantages: (1) Gaussian splats allow detections from multiple vendors to be combined in a statistically meaningful manner through aggregation of their spatial statistics, and (2) the sufficient statistics of a Gaussian distribution can be expressed as linear moment sums, enabling secure fusion through homomorphic addition without exposing the underlying detections. Using these two properties \textsf{Sarus} transforms perception fusion into a linear aggregation problem, which can be performed directly on encrypted data while preserving the information required to reconstruct fused object detections.

A detection produced by an object detector is represented by an
axis-aligned bounding box $\mathbf{b} = [x_1,y_1,x_2,y_2]$. Geometrically, the bounding box defines a spatial region as follows: $$\mathbf{b} = \{(x,y) \mid x_1 \le x \le x_2,\; y_1 \le y \le y_2\}.$$

Let the width and height of the bounding box be $w = x_2 - x_1$, and $h = y_2 - y_1$ respectively and its geometric center be $c_x = \frac{x_1+x_2}{2}$ and $c_y = \frac{y_1+y_2}{2}$ respectively. Since the exact object center within the detected box is uncertain, we model spatial uncertainty by assuming that the true object center is uniformly distributed within the bounding box. This assumption should be interpreted as a local non-informative prior over the admissible support region, not as a uniform distribution over the entire scene. In particular, the support is constrained by the detected bounding box, the image plane, and the physically feasible region associated with the candidate object; therefore, physically invalid locations are excluded~\cite{elfes1989occupancy,phan2018calibrating,wang2020inferring}. Let $X$ and $Y$ denote the horizontal and vertical object coordinates, respectively. Then under uniform uncertainty assumption, we have: 
$$X \sim \mathrm{Uniform}(x_1,x_2), \qquad Y \sim \mathrm{Uniform}(y_1,y_2).$$

The corresponding probability density functions are given by:
\[
f_X(x)=
\begin{cases}
\frac{1}{w}, & x_1 \le x \le x_2 \\
0, & \text{otherwise}
\end{cases}
\qquad
f_Y(y)=
\begin{cases}
\frac{1}{h}, & y_1 \le y \le y_2 \\
0, & \text{otherwise}.
\end{cases}
\]

Assuming independence between the two spatial coordinates,
the joint density inside the bounding box becomes

\[
f_{\mathbf{b}}(x,y)=
\begin{cases}
\frac{1}{wh}, & (x,y)\in \mathbf{b} \\
0, & \text{otherwise}.
\end{cases}
\]

\begin{figure}[!t]
     \centering
     \begin{subfigure}[b]{0.49\textwidth}
         \centering
         \includegraphics[width=\textwidth]{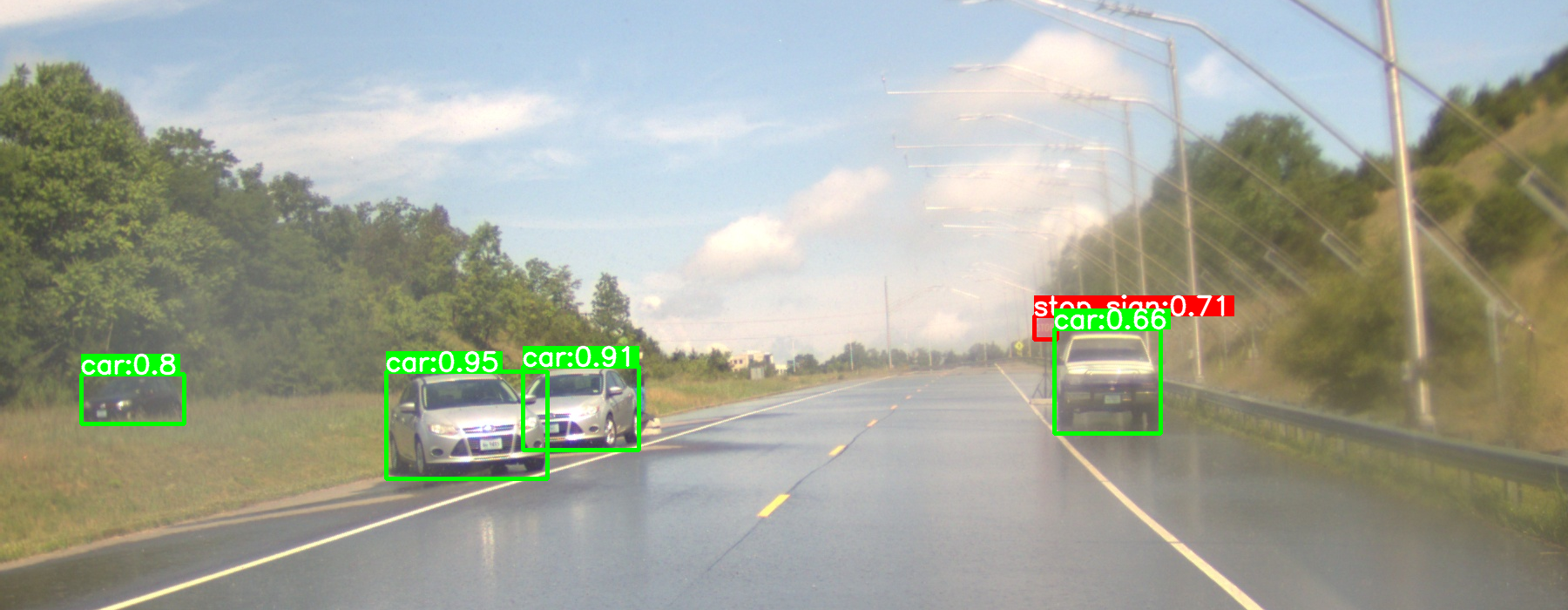}
         \caption{Stop sign detected $ 59.46m$, pedestrian undetected $47.07m$.}
         \label{fig-rt-detr-detection}
     \end{subfigure}
     %\hfill
     \begin{subfigure}[b]{0.49\textwidth}
         \centering
         \includegraphics[width=\textwidth]{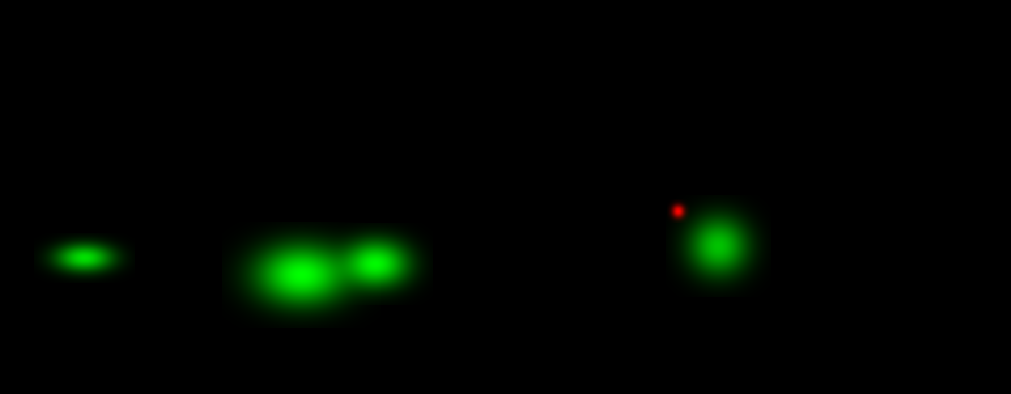}
         \caption{Gaussian Splat for the detection.}
         \label{fig-rt-detr-splat}
     \end{subfigure}
    \caption{RT-DETR detection.}
    \label{fig-rt-detr-detection-splat}
\end{figure}

\begin{figure}[!t]
     \centering
     \begin{subfigure}[b]{0.49\textwidth}
         \centering
         \includegraphics[width=\textwidth]{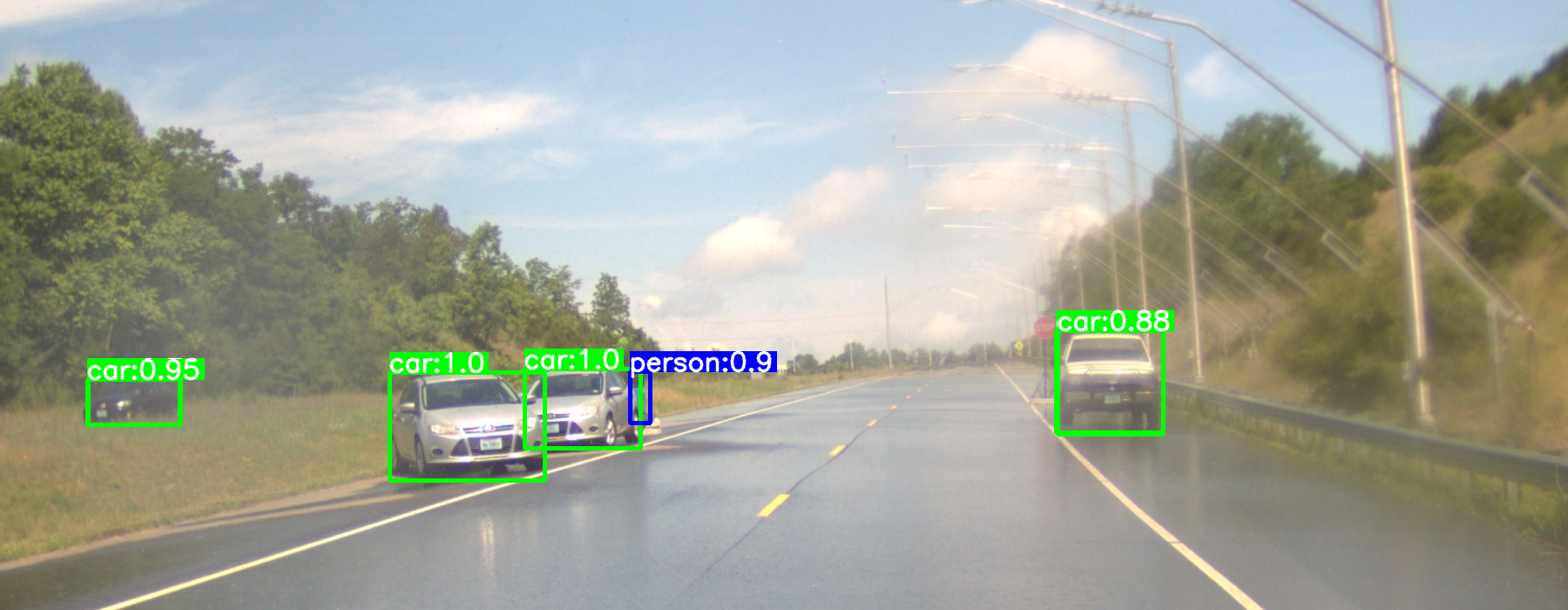}
         \caption{Stop sign undetected $ 59.46m$, pedestrian detected $47.07m$.}
         \label{fig-detr101-detection}
     \end{subfigure}
     %\hfill
     \begin{subfigure}[b]{0.49\textwidth}
         \centering
         \includegraphics[width=\textwidth]{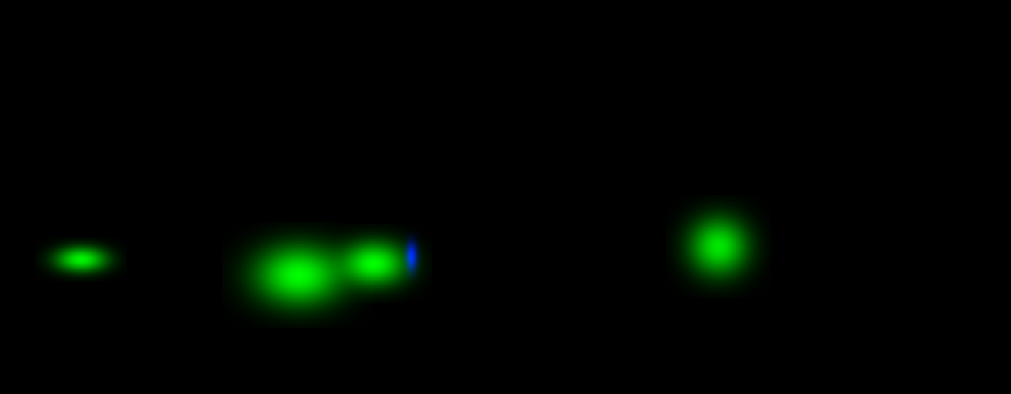}
         \caption{Gaussian Splat for the detection.}
         \label{fig-detr101-splat}
     \end{subfigure}
    \caption{DETR101 detection.}
    \label{fig-detr101-detection-splat}
\end{figure}

The mean and variance of the induced spatial distribution are therefore: $\mu_x = \mathbb{E}[X] = c_x$, $\mu_y = \mathbb{E}[Y] = c_y$ and $\sigma_x^2 = \mathrm{Var}(X)=\frac{w^2}{12}$ and
$\sigma_y^2 = \mathrm{Var}(Y)=\frac{h^2}{12}$. While the uniform model captures the spatial extent of the bounding box, \textsf{Sarus} represents detections using a Gaussian splat whose parameters preserve the mean of the distribution while allowing a tunable spatial spread. This spacial spread is controlled by a scaling parameter $\kappa > 0$, hence:

\begin{equation}
    \label{eq:kappa-variance}
    \sigma_x^2 = \left(\kappa \frac{w}{2}\right)^2, \qquad
    \sigma_y^2 = \left(\kappa \frac{h}{2}\right)^2.
\end{equation}
Note, when $\kappa = \frac{1}{\sqrt{3}}$, equation~\eqref{eq:kappa-variance} matches the variance induced by the uniform spatial uncertainty of the bounding box. Under this representation, each detection is approximated by the Gaussian
distribution: 
\begin{equation}
    \label{eq: gaussian-distribution}
    \mathcal{N}\big((\mu_x,\mu_y),
\mathrm{diag}(\sigma_x^2,\sigma_y^2)\big),
\end{equation}
which we refer to as a \emph{Gaussian splat}: a continuous spatial model of the detection preserving the geometric information implied by the bounding box and used for encrypted moment aggregation. Figure~\ref{fig-rt-detr-detection-splat} and Figure~\ref{fig-detr101-detection-splat} show detection in the image space (Figure~\ref{fig-rt-detr-detection} and Figure~\ref{fig-detr101-detection}) and their respective gaussian splats (Figure~\ref{fig-rt-detr-splat} and Figure~\ref{fig-detr101-splat}).


\paragraph{Moment-Based Detection Encoding:} Let a detection $\mathbf{b}$ with class confidence $p$ be modeled by the Gaussian distribution of equation~\eqref{eq: gaussian-distribution}. Let $\alpha$ be the trust weight associated with some vendor $v \in \mathcal{V}$, such that the effective contribution weight of
the detection is defined as: $w = \alpha p$. Then \textsf{Sarus} encodes each detection using a vector of weighted spatial moments:
\begin{equation}
    \label{eq:wt-moment-vector}
    \mathbf{m} = w \cdot \begin{bmatrix}
        1,\; \mu_x,\; \mu_x^2,\; \sigma_x^2,\; \mu_y,\; \mu_y^2,\; \sigma_y^2
    \end{bmatrix}.
\end{equation}
Each component the vector in equation~\eqref{eq:wt-moment-vector} represents a sufficient statistic of the spatial distribution and is linear with respect to aggregation across detections. Consequently, moment vectors from multiple detections can be accumulated through simple summation.

\paragraph{Aggregated Moments.}
Let $\mathcal{H}$ denote an object hypothesis, defined as a set of
detections that are hypothesized to correspond to the same underlying object instance. Formally, $\mathcal{H}$ is a subset of detections whose spatial coordinates and class predictions are mutually consistent under the fusion model. Let $\mathbf{m}_k$ denote the moment vector produced by the $k^{\text{th}}$ detection in this set, where $k \in \mathcal{H}$. The aggregated statistics of the hypothesis are obtained by: $$\mathbf{s} = \sum_{k \in \mathcal{H}} \mathbf{m}_k.$$

Expanding the vector yields

\begin{equation}
\label{eq:aggregated-wt-moment-vector}
    \mathbf{s} =
    \begin{bmatrix}
    S_w,\;
    S_{w\mu_x},\;
    S_{w\mu_x^2},\;
    S_{w\sigma_x^2},\;
    S_{w\mu_y},\;
    S_{w\mu_y^2},\;
    S_{w\sigma_y^2}
    \end{bmatrix}.
\end{equation}

These aggregated quantities correspond to the sufficient statistics of the Gaussian representation associated with the hypothesis $\mathcal{H}$.

\begin{lemma}[Linearity and Sufficiency]
\label{lem:moment-linearity}
Let each contributing detection $k \in K$ produce a moment vector
\[
\mathbf{m}_k =
\big[
w_k,\;
w_k\mu_{x,k},\;
w_k\mu_{x,k}^2,\;
w_k\sigma_{x,k}^2,\;
w_k\mu_{y,k},\;
w_k\mu_{y,k}^2,\;
w_k\sigma_{y,k}^2
\big],
\]
where $w_k=\alpha_{v(k)}p_k$ is the weighted confidence of the detection for vendor $v \in \mathcal{V}$, such that these moment vectors are aggregated through summation to obtain
\[
\mathbf{s}=\sum_k \mathbf{m}_k
=
[S_w,\;S_{w\mu_x},\;S_{w\mu_x^2},\;S_{w\sigma_x^2},\;
 S_{w\mu_y},\;S_{w\mu_y^2},\;S_{w\sigma_y^2} ].
\]

Then:

\begin{enumerate}
\item \textbf{(Linearity)}  
The aggregated moment vector $\mathbf{s}$ can be computed using only
component-wise additions. Consequently, if the vectors $\mathbf{m}_k$ are
encrypted, homomorphic addition yields the exact plaintext sums after
decryption.

\item \textbf{(Sufficiency)}  
The fused Gaussian parameters $(\mu_x,\mu_y,\sigma_x^2,\sigma_y^2)$ can be recovered from $\mathbf{s}$ as

\[
\mu_x=\frac{S_{w\mu_x}}{S_w},
\qquad
\mu_y=\frac{S_{w\mu_y}}{S_w},
\]

\[
\sigma_x^2=
\frac{S_{w\sigma_x^2}+S_{w\mu_x^2}}{S_w}-\mu_x^2,
\qquad
\sigma_y^2=
\frac{S_{w\sigma_y^2}+S_{w\mu_y^2}}{S_w}-\mu_y^2.
\]

Hence, the 7 aggregated moments are sufficient to reconstruct the fused Gaussian representation of the detections.
\end{enumerate}
\end{lemma}

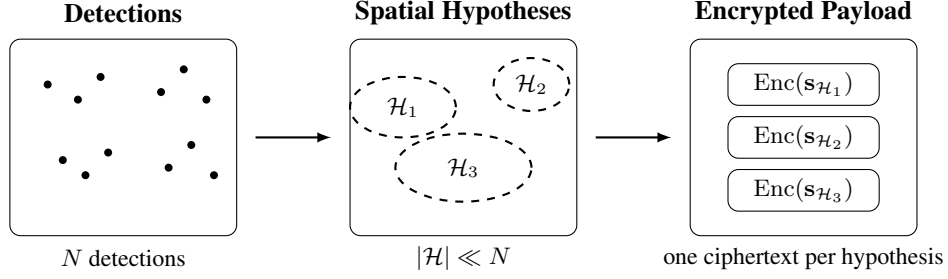
\begin{figure}[t]
\centering
\begin{tikzpicture}[>=latex, font=\small]

% --- Left panel: detections ---
\draw[rounded corners] (0,0) rectangle (3,2.6);
\node[font=\bfseries] at (1.5,2.95) {Detections};

% Many detections
\fill (0.5,2.0) circle (1.5pt);
\fill (0.9,1.8) circle (1.5pt);
\fill (1.2,2.1) circle (1.5pt);
\fill (2.0,1.9) circle (1.5pt);
\fill (2.3,2.2) circle (1.5pt);
\fill (2.6,1.8) circle (1.5pt);

\fill (0.7,1.0) circle (1.5pt);
\fill (1.0,0.8) circle (1.5pt);
\fill (1.3,1.1) circle (1.5pt);
\fill (2.1,0.9) circle (1.5pt);
\fill (2.4,1.2) circle (1.5pt);
\fill (2.7,0.8) circle (1.5pt);

\node at (1.5,-0.3) {$N$ detections};

\draw[->, thick] (3.25,1.3) -- (4.25,1.3);

\draw[rounded corners] (4.5,0) rectangle (7.5,2.6);
\node[font=\bfseries] at (6.0,2.95) {Spatial Hypotheses};

\draw[thick, dashed] (5.2,1.7) ellipse (0.7 and 0.4);
\draw[thick, dashed] (6.9,2.0) ellipse (0.5 and 0.35);
\draw[thick, dashed] (6.0,0.9) ellipse (0.9 and 0.45);

% Labels => fix later
\node at (5.2,1.7) {$\mathcal{H}_1$};
\node at (6.9,2.0) {$\mathcal{H}_2$};
\node at (6.0,0.9) {$\mathcal{H}_3$};

\node at (6.0,-0.3) {$|\mathcal{H}| \ll N$};

\draw[->, thick] (7.75,1.3) -- (8.75,1.3);

\draw[rounded corners] (9.0,0) rectangle (12.0,2.6);
\node[font=\bfseries] at (10.5,2.95) {Encrypted Payload};

\node[draw, rounded corners, minimum width=2.0cm, minimum height=0.45cm] at (10.5,2.0) {$\mathrm{Enc}(\mathbf{s}_{\mathcal{H}_1})$};
\node[draw, rounded corners, minimum width=2.0cm, minimum height=0.45cm] at (10.5,1.3) {$\mathrm{Enc}(\mathbf{s}_{\mathcal{H}_2})$};
\node[draw, rounded corners, minimum width=2.0cm, minimum height=0.45cm] at (10.5,0.6) {$\mathrm{Enc}(\mathbf{s}_{\mathcal{H}_3})$};

\node at (10.5,-0.3) {one ciphertext per hypothesis};

\end{tikzpicture}
\caption{Illustration of vendor-side moment aggregation in \textsf{Sarus}. A large set of detections is first grouped into a smaller number of spatial object hypotheses, and only one encrypted aggregated moment vector is transmitted per hypothesis.}
\label{fig:vendor-side-aggregation}
\end{figure}

The design choice of local aggregation of moment contributions at the vendor before encryption provides the following advantages:
\begin{enumerate}
    \item \textbf{Reduced encrypted payload size:} Without aggregation, each detection may contribute up to four encrypted moment vectors (due to bilinear assignment, refer section~\ref{sec:sub:spatial-bin-bilinear-assignment}), resulting in $O(N)$ encrypted messages for $N$ detections. With aggregation, the payload scales with the number of object hypotheses $|\mathcal{H}|$, typically $|\mathcal{H}| \ll N$ (see Figure~\ref{fig:vendor-side-aggregation}). For example, if a vendor produces $N=200$ detections, the naive approach may require up to $800$ encrypted vectors. In practice many detections fall within the same spatial hypothesis region and therefore contribute to a common aggregated moment vector, considerably reducing the number of encrypted vectors (e.g., $|\mathcal{H}| \approx 30$ hypotheses.)
    \item \textbf{Lower encryption cost:} Encryption is performed once per hypothesis rather than once per detection contribution, reducing the number of cryptographic operations.
    \item \textbf{Improved network scalability:} Across multiple vendors the reduction becomes substantial. For instance, with $50$ vendors each producing $200$ detections, the naive approach would transmit roughly $50 \times 800 \approx 40000$ ciphertexts, whereas aggregation reduces this to approximately $50 \times 30 \approx 1500$ ciphertexts.
    \item \textbf{Efficient homomorphic fusion:} Because moment aggregation is linear (Lemma~\ref{lem:moment-linearity}), encrypted hypothesis statistics can be combined across vendors using homomorphic addition.
\end{enumerate}

We now introduce the spatial binning and soft assignment procedure, which defines the sets of detections that contribute to each object hypothesis $\mathcal{H}$. Detections are associated with nearby grid regions based on the location of their Gaussian centers, and their moment vectors are distributed to neighboring bins using bilinear weights. The resulting moment contributions are then aggregated according to Eq.~\ref{eq:aggregated-wt-moment-vector} leveraging the linearity of the moment representation.

\begin{algorithm}[t]
\caption{Class-Specific Spatial Grid Generation}
\label{alg:spatial-grid-generation}
    \begin{algorithmic}[1]
    \Require List of classes: $C$, List of spatial anchors for each class: $S$, List of strides for each class: $s$, Image width: $W$, Image height: $H$.
    \State $G \leftarrow \phi$ \Comment{\textsl{Initialize grid for each class $c$ in $C$}}
    \For{$c=1$ to $|C|$}
        \State $G[c] \leftarrow \phi$ \Comment{\textsl{Initialize $G$ for current class $c$}}
        \State $n_x \leftarrow \lceil \frac{W - \frac{S[c]}{2}}{s[c]} \rceil$, $n_y \leftarrow \lceil \frac{H - \frac{S[c]}{2}}{s[c]} \rceil$ \Comment{\textsl{Number of bins for class $c \in C$}}
        \For{$j=1$ to $n_y$}
            \State $y = \frac{S[c]}{2} + (j-1)\cdot s[c]$
            \For{$i=1$ to $n_x$}
                 \State $x = \frac{S[c]}{2} +  (i-1)\cdot s[c]$
                \State  $G[c].\text{append}((x,y))$ \Comment{\textsl{Append grid center for each bin}}
            \EndFor
        \EndFor
    \EndFor
    \State \textbf{Return} $G$
    \end{algorithmic}
\end{algorithm}

\begin{figure}[!t]
    \centering
    \includegraphics[width=1.0\linewidth]{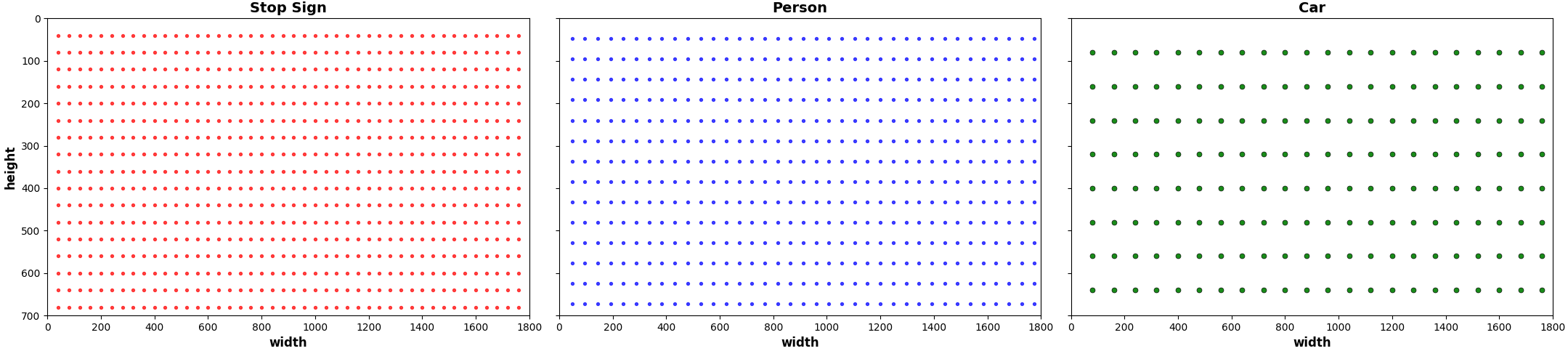}
    \caption{Bin centers using spatial anchors for image with width: $1800$ and height $700$. Stop sign: $748$ bins $(S=80, s=40)$, Person: $518$ bins $(S=96, s=48)$ and Car: $176$ bins $(S=160, s=80)$.}
    \label{fig:grid-generation}
\end{figure}

\subsubsection{Spatial Binning and Soft Assignment}
\label{sec:sub:spatial-bin-bilinear-assignment}
Direct aggregation of detections across the entire image would allow
spatially distant objects to influence each other, potentially producing unstable fusion results. To ensure that only spatially consistent detections are combined, the image is partitioned into a
class-specific spatial grid and aggregation is performed independently within each grid cell. Let $S_c$ and $s_c$ be spacial anchor and stride for a given class $c$, then Algorithm~\ref{alg:spatial-grid-generation} creates such a grid. Since grid creation can be precomputed for a given image size and class, we safely ignore Algorithm~\ref{alg:spatial-grid-generation} in complexity calculation in section~\ref{sec:vendor-payload-complexity}.

\paragraph{Soft Bin Assignment:}
Let $(\mu_x,\mu_y)$ denote the center of the Gaussian splat derived
from a bounding box. Then instead of assigning the detection to a single bin, its contribution is distributed to the four neighboring bins using bilinear interpolation. Let $(x_i,y_j)$ denote the center of bin $(i,j)$. Then fractional offsets of the detection center within the local grid cell are: $t_x = \frac{\mu_x - x_i}{s_c}$ and $t_y = \frac{\mu_y - y_j}{s_c}$ and the four neighboring bins are: $\{(i,j), (i+1,j), (i,j+1), (i+1,j+1)\}$ with respective bilinear interpolation weights given by: $$
\omega_{00} = (1-t_x)(1-t_y),\; \omega_{10} = t_x(1-t_y), \;
\omega_{01} = (1-t_x)t_y, \; \omega_{11} = t_x t_y, $$
where, $\omega_{00} + \omega_{10} + \omega_{01} + \omega_{11} = 1$, such that the total contribution of the detection is preserved.

\paragraph{Moment Accumulation.}
For each bin receiving a contribution, the weighted moment vector $\mathbf{m}_k$ (introduced earlier) is accumulated as: $\mathbf{s}_{c,(i,j)} \leftarrow \mathbf{s}_{c,(i,j)} + \omega_{k \rightarrow (i,j)}\,\mathbf{m}_k$, where $(i,j)$ is the recieving bin. The same update is applied for the neighboring bins i.e., $(i+1,j)$, $(i,j+1)$, and $(i+1,j+1)$ using their corresponding
weights. Since each detection contributes to at most four bins, the update cost per detection is constant. Consequently, the construction of per-class-bin moment accumulators scales linearly with the number of detections produced by a vendor and preserving smooth spatial transitions across bin boundaries, which improves the stability of the fused detections.

\subsection{Vendor Payload Construction}
Let vendor $v \in \mathcal{V}$ produce a set of detections $\mathcal{D}_v$ such that: $\mathcal{D}_v = \{ d_{v,1}, d_{v,2}, \dots, d_{v,N_v} \}$, where $N_v$ is the total number of detections produced by $v$. Each detection $d_{v,k}$, where $k \in \{1, \ldots,  N_v\}$, is first converted into its Gaussian splat representation as described in Section~\ref{sec:detection-to-gaussian-splat}, yielding parameters $(\mu_{x,v,k}, \mu_{y,v,k}, \sigma_{x,v,k}, \sigma_{y,v,k})$ for the $k^{\text{th}}$ detection. Let $p_{v,k}$ be the detection confidence for the $k^{\text{th}}$ detection and the vendor trust weight $\alpha_v$, then the detection weight is defined as: $w_{v,k} = \alpha_v p_{v,k}$. The detection is then encoded as the moment vector as follows:
$$\mathbf{m}_{v,k}
=
w_{v,k}
\begin{bmatrix}
1,\;
\mu_{x,v,k},\;
\mu_{x,v,k}^2,\;
\sigma_{x,v,k}^2,\;
\mu_{y,v,k},\;
\mu_{y,v,k}^2,\;
\sigma_{y,v,k}^2
\end{bmatrix}.$$
Using the soft assignment procedure described previously, detection
$d_{v,k}$ contributes to at most four neighboring bins with weights
$\omega_{k\rightarrow(i,j)}$. The vendor accumulates these
contributions locally:
$$\mathbf{s}_{v,c,(i,j)}
\leftarrow
\mathbf{s}_{v,c,(i,j)}
+
\omega_{k\rightarrow(i,j)} \, \mathbf{m}_{v,k}.$$
This accumulation is performed independently for each object class
$c$ and spatial bin $(i,j)$. After processing all detections, vendor $v$ obtains a set of per-bin moment accumulators as follows:
$$\mathbf{s}_{v,c,(i,j)}
=
\begin{bmatrix}
S_w,\;
S_{w\mu_x},\;
S_{w\mu_x^2},\;
S_{w\sigma_x^2},\;
S_{w\mu_y},\;
S_{w\mu_y^2},\;
S_{w\sigma_y^2}
\end{bmatrix}.$$
These aggregated statistics constitute sufficient information to
reconstruct the fused Gaussian parameters after aggregation.  
\paragraph{Payload Structure:}
After local aggregation, vendor $v$ constructs a payload indexed by the class-bin key i.e., $\texttt{key} \leftarrow (c, (i, j))$. Let $\mathsf{Enc}_{\mathsf{HE}}(\cdot)$ denote homomorphic encryption scheme which supports linear operation in the encrypted domain (\textsf{Sarus} uses CKKS\cite{cheon2017homomorphic}-based homomorphic encryption scheme that provides additive homomorphism). To enable privacy-preserving multi-vendor fusion, each moment vector
is encrypted using $\mathsf{Enc}_{\mathsf{HE}}(\cdot)$. For each occupied $\texttt{key}$, the vendor stores three quantities:
\begin{equation}
    \label{eq:vendor-payload}
    \begin{split}
        \mathbf{c}_{v,\texttt{key}} &= \mathsf{Enc}_{\mathsf{HE}}(\mathbf{s}_{v, \texttt{key}}), \\
        M_{v, \texttt{key}} &= \sum_{k \in \mathcal{H}_{v,\texttt{key}}} \omega_{k\rightarrow \texttt{key}}\,(\alpha_v p_{v, \texttt{key}}), \\
        C_{v,\texttt{key}} &= \sum_{k \in \mathcal{H}_{v,\texttt{key}}} \omega_{k\rightarrow \texttt{key}}\,\alpha_v,
    \end{split}
\end{equation}
where, $\mathcal{H}_{v,\texttt{key}}$ denotes the set of detections from $v$ whose soft assignment contributes to $\texttt{key}$, and $\omega_{k\rightarrow \texttt{key}}$ is the corresponding bilinear assignment weight. The quantity $M_{v, \texttt{key}}$ represents the aggregated weighted confidence mass for $\texttt{key}$ and $C_{v,\texttt{key}}$ represents the aggregated weight count. The  payload is the set defined as: $$\mathcal{P}_v = \left\{
\bigl(\mathbf{c}_{v,\texttt{key}},M_{v,\texttt{key}},C_{v,\texttt{key}}\bigr) \;\middle|\; \texttt{key}=(c,(i,j)) \right\},$$ for all occupied class-bin keys. Algorithm~\ref{alg:he-vendor-payload-construction} presents the complete pseudocode that a vendor $v \in \mathcal{V}$ uses to generate their respective payloads.

\begin{algorithm}[!t]
\caption{Vendor Payload Construction}
\label{alg:he-vendor-payload-construction}
    \begin{algorithmic}[1]
        \Require Detection: $\mathcal{D}$, $\texttt{image\_width}: W$, $\texttt{image\_height}: H$
        \Ensure 
            \parbox[t]{0.9\linewidth}{
            Each $\texttt{detection} \in \mathcal{D}$:\\
            \quad $\texttt{detection}$: $\{\texttt{probability}, \texttt{class\_id}, \texttt{bbox}:[x_1, y_1,x_2,y_2], \texttt{spacial\_anchor}: S, \texttt{stride}: s\}$
        }
        \For{each detection $d \in \mathcal{D}$}
            \State $p \leftarrow d.\texttt{get}(\texttt{probability})$, $\texttt{cls} \leftarrow d.\texttt{get}(\texttt{class\_id})$, $\mathbf{b} \leftarrow d.\texttt{get}(\texttt{bbox})$
            \State $(c_x, c_y, \sigma^2_x, \sigma^2_y) \leftarrow \textbf{BoxToGaussianParams}(\mathbf{b}, \kappa)$
            \State $\mathbf{m} \leftarrow [\alpha p, \alpha c_x, \alpha c_x^2, \alpha \sigma^2_x, \alpha c_y, \alpha c_y^2, \alpha \sigma^2_y]$
            \State $\mathcal{B} \leftarrow \textbf{SoftAssignBins}(c_x, c_y, d.\text{get}(\texttt{spacial\_anchor}), d.\text{get}(\texttt{stride}), W, H)$
            \For{each $(\texttt{bin},\omega) \in \mathcal{B}$}
                \State $\texttt{key} \leftarrow (\texttt{cls}, \texttt{bin\_id})$
                \State $\texttt{accumulator}[\texttt{key}] \leftarrow \texttt{accumulator}[\texttt{key}] + \omega\cdot \mathbf{m}$
                \State $\texttt{mass\_by\_key}[\texttt{key}] \leftarrow \texttt{mass\_by\_key}[\texttt{key}] + \omega \cdot (\alpha p)$
                \State $\texttt{count\_by\_key}[\texttt{key}] \leftarrow \texttt{count\_by\_key}[\texttt{key}] + \omega \cdot \alpha$
            \EndFor
        \EndFor
        \State $\texttt{cipher\_by\_key} \leftarrow \emptyset$
        \For{each $\texttt{key} \in \mathrm{keys}(\texttt{accumulator})$}
            \State $\texttt{ciphertext\_bytes} \leftarrow \textrm{Enc}_{\mathrm{HE}}(\texttt{accumulator}[\texttt{key}])$ \Comment{Homomorphic encryption}
            \State $\texttt{cipher\_by\_key}[\texttt{key}] \leftarrow \texttt{ciphertext\_bytes}$
        \EndFor
        \State \textbf{Return} $(\texttt{cipher\_by\_key}, \texttt{mass\_by\_key}, \texttt{count\_by\_key})$
    \end{algorithmic}
    \hfill\\
    \textbf{BoxToGaussianParams}$(\mathbf{b}, \kappa)$\textbf{:}
    \begin{algorithmic}[1]
        \State $x_1, y_1, x_2, y_2 \leftarrow \mathbf{b}$
        \State $w \leftarrow (x_2 - x_1)$, $h \leftarrow (y_2 - y_1)$
        \State $c_x \leftarrow \frac{x_1 + x_2}{2}$, $c_y \leftarrow \frac{y_1 + y_2}{2}$
        \State $\sigma^2_x \leftarrow \left(\kappa \cdot \frac{w}{2} \right)^2$, $\sigma^2_y \leftarrow \left(\kappa \cdot \frac{h}{2} \right)^2$
        \State \textbf{Return} $(c_x, c_y, \sigma^2_x, \sigma^2_y)$
    \end{algorithmic}
    \hfill\\
    \textbf{SoftAssignBins}$(\mu_x, \mu_y, S, s, W, H)$\textbf{:}
    \begin{algorithmic}[1]
        \State $(n_x, n_y, \_) \leftarrow \textbf{SpatialGridGenerator}(W, H, S, s)$
        \State $i \leftarrow \left\lfloor (\mu_x - S/2)/s \right\rfloor + 1$,$j \leftarrow \left\lfloor (\mu_y - S/2)/s \right\rfloor + 1$
        \State $i \gets \max(1,\min(i,n_x-1))$, \;\; $j \gets \max(1,\min(j,n_y-1))$
    
        \State $x_i \gets S/2 + (i-1)s$, \;\; $y_j \gets S/2 + (j-1)s$
        \State $t_x \gets (\mu_x - x_i)/s$, \;\; $t_y \gets (\mu_y - y_j)/s$
        \State $t_x \gets \min(1,\max(0,t_x))$, \;\; $t_y \gets \min(1,\max(0,t_y))$
    
        \State $\mathcal{B} \gets \emptyset$
        \State $b_{00}\gets(i,j)$,     \;\; $\omega_{00}\gets(1-t_x)(1-t_y)$
        \State $b_{10}\gets(i+1,j)$,   \;\; $\omega_{10}\gets t_x(1-t_y)$
        \State $b_{01}\gets(i,j+1)$,   \;\; $\omega_{01}\gets(1-t_x)t_y$
        \State $b_{11}\gets(i+1,j+1)$, \;\; $\omega_{11}\gets t_x t_y$
    
        \For{each $(\texttt{bin\_id},\omega)\in\{(b_{00},\omega_{00}),(b_{10},\omega_{10}),(b_{01},\omega_{01}),(b_{11},\omega_{11})\}$}
            \If{$\omega>0$}
                \State add $(\texttt{bin\_id},\omega)$ to $\mathcal{B}$
            \EndIf
        \EndFor
    
        \State $Z \gets \sum_{(\texttt{bin\_id},\omega)\in\mathcal{B}} \omega$
        \If{$Z>0$}
            \For{each $(\texttt{bin\_id},\omega)\in\mathcal{B}$}
                \State $\omega \gets \omega/Z$
            \EndFor
        \EndIf
    
        \State \Return $\mathcal{B}$
    \end{algorithmic}
\end{algorithm}

\subsection{Encrypted Multi-Vendor Fusion by Server}
Let $\mathcal{V}$ denote the set of participating vendors such that each vendor $v \in \mathcal{V}$ transmits a payload, $\mathcal{P}_v = \left\{(\mathbf{c}_{v,\texttt{key}},M_{v,\texttt{key}},C_{v,\texttt{key}})\right\}$
where $\texttt{key}=(c,(i,j))$ denotes the class-bin key, $\mathbf{c}_{v,\texttt{key}}$ is the encrypted aggregated moment vector, and $M_{v,\kappa}$ and $C_{v,\texttt{key}}$ are the associated statistics defined above. The fusion server first computes the union of all keys contributed by vendors i.e., $$\mathcal{K} = \bigcup_{v \in \mathcal{V}} \mathrm{keys}(\mathcal{P}_v),$$
where, each $\texttt{key} \in \mathcal{K}$ corresponds to a spatial
hypothesis.

\paragraph{Homomorphic Moment Fusion:}
For each $\texttt{key}$, the server aggregates encrypted moment vectors across vendors using the additive homomorphism of the encryption scheme: 
\begin{equation}
    \label{eq:server-homomorphic-addition}
    \mathbf{c}_{\texttt{key}}^{\,\mathrm{fused}} = \bigoplus_{v \in \mathcal{V}_\texttt{key}}
\mathbf{c}_{v,\texttt{key}}
\end{equation}
where $\mathcal{V}_\texttt{key} \subseteq \mathcal{V}$ denotes the set of vendors that contributed to \texttt{key}, and $\oplus$ denotes homomorphic ciphertext addition. Since the moment representation is linear (refer Lemma~\ref{lem:moment-linearity}), hence, the operation in equation~\eqref{eq:server-homomorphic-addition} produces an encryption of the fused moment vector:
$$
\mathbf{c}_{\texttt{key}}^{\,\mathrm{fused}}
=
\mathsf{Enc}\!\left(
\sum_{v \in \mathcal{V}_\texttt{key}}
\mathbf{s}_{v,\texttt{key}}
\right).
$$
The accompanying statistics are aggregated as follows:
$$ M_{\texttt{key}}^{\mathrm{fused}}
=
\sum_{v \in \mathcal{V}_\texttt{key}} M_{v,\texttt{key}},
\qquad
C_{\texttt{key}}^{\mathrm{fused}}
=
\sum_{v \in \mathcal{V}_\texttt{key}} C_{v,\texttt{key}}. $$

\paragraph{Fused Payload:}
The resulting fused representation maintained by the server is given by the following set:$$\mathcal{P}^{\mathrm{fused}}
=
\left\{(
\mathbf{c}_{\texttt{key}}^{\mathrm{fused}},
M_{\texttt{key}}^{\mathrm{fused}},
C_{\texttt{key}}^{\mathrm{fused}})
 \;\middle|\; \texttt{key}=(c,(i,j)) \right\},$$
which contains encrypted fused moment vectors and the associated  statistics for every occupied \texttt{key}. Algorithm~\ref{alg:he-fusion-by-server} presents the pseudocode that the server uses to create fused payload.

\begin{algorithm}[!t]
\caption{HE-Encrypted Multi-Vendor Fusion by Server}
\label{alg:he-fusion-by-server}
    \begin{algorithmic}[1]
        \Require Vendor Payloads $\mathcal{V} =\{p_1,\dots,p_{V}\}$ where $|\mathcal{V}| = V$
        \Ensure 
            \parbox[t]{0.9\linewidth}{
            Each vendor payload $p_v$, $\forall v \in \mathcal{V}$:\\
            \quad $\texttt{cipher\_by\_key}$: $\{\texttt{key}:(\texttt{class\_id} , \texttt{bin\_id)}, \texttt{ciphertext\_bytes}\}$ \Comment{HE-encrypted $\mathbf{v}$}\\
            \quad $\texttt{mass\_by\_key}$: $\{\texttt{key}:(\texttt{class\_id} , \texttt{bin\_id)}, \texttt{float}\}$ \Comment{Confidence mass}\\
            \quad $\texttt{count\_by\_key}$: $\{\texttt{key}:(\texttt{class\_id} , \texttt{bin\_id)}, \texttt{float}\}$ \Comment{Detection count}
        }
        \State $\texttt{cipher\_list} \leftarrow \emptyset$, $\texttt{mass\_list} \leftarrow \emptyset$, $\texttt{count\_list} \leftarrow \emptyset$, $\texttt{all\_keys} \leftarrow \emptyset$ \State $\texttt{fused\_cipher\_by\_key} \leftarrow \emptyset$, $\texttt{fused\_mass\_by\_key} \leftarrow \emptyset$, $\texttt{fused\_count\_by\_key} \leftarrow \emptyset$
        \For{$v \in \mathcal{V}$}
            \State $\texttt{cipher\_list}\text{.append}(p_v[\texttt{cipher\_by\_key}])$  
            \State $\texttt{mass\_list}\text{.append}(p_v[\texttt{mass\_by\_key}])$, 
            \State $\texttt{count\_list}\text{.append}(p_v[\texttt{count\_by\_key}])$ 
        \EndFor
        \For{each $\texttt{cipher} \in \texttt{cipher\_list}$}
            \State $\texttt{all\_keys} \leftarrow 
            \texttt{all\_keys} \cup \mathrm{keys}(\texttt{cipher})$ \Comment{Collect all unique $(\texttt{class\_id},\texttt{bin\_id})$ keys across vendors payloads}
        \EndFor
        
        \If{$\texttt{all\_keys} = \emptyset$}
            \State \textbf{raise error} \texttt{"No ciphertexts available"}
        \EndIf

        \For{each $\texttt{key} \in \texttt{all\_keys}$}
            \State $\texttt{fused\_cipher\_by\_key}[\texttt{key}] \leftarrow \perp$, $\texttt{fused\_mass\_by\_key}[\texttt{key}] \leftarrow 0$, $\texttt{fused\_count\_by\_key}[\texttt{key}] \leftarrow 0$ 
            \State $\texttt{accumulator} \leftarrow \texttt{None}$
            \For{each $\texttt{cipher} \in \texttt{cipher\_list}$}
                \State $c \leftarrow \texttt{cipher}.\text{get}(\texttt{key})$
                \If{$c \neq \texttt{None}$}
                    \If{\texttt{accumulator} $=$ \texttt{None}}
                        \State $\texttt{accumulator} \leftarrow c$
                    \Else
                        \State $\texttt{accumulator} \leftarrow \texttt{accumulator} + c$ \Comment{HE addition based on key $(\texttt{class\_id},\texttt{bin\_id})$}
                    \EndIf
                \EndIf
            \EndFor
            \State $\texttt{fused\_cipher\_by\_key}[\texttt{key}] \leftarrow \texttt{accumulator}$
            \For{each $\texttt{mass} \in \texttt{mass\_list}$}
                \State $\texttt{fused\_mass\_by\_key}[\texttt{key}] \leftarrow \texttt{fused\_mass\_by\_key}[\texttt{key}] + \texttt{mass}.\text{get}(\texttt{key})$
            \EndFor
            \For{each $\texttt{count} \in \texttt{count\_list}$}
                \State $\texttt{fused\_count\_by\_key}[\texttt{key}] \leftarrow \texttt{fused\_count\_by\_key}[\texttt{key}] + \texttt{count}.\text{get}(\texttt{key})$
            \EndFor
        \EndFor
        \State \textbf{Return} $(\texttt{fused\_cipher\_by\_key}, \texttt{fused\_mass\_by\_key}, \texttt{fused\_count\_by\_key})$    
    \end{algorithmic}
\end{algorithm}

\subsection{Fused Payload to Detection Reconstruction}
Vendor $v \in \mathcal{V}$ operates on the fused payload shared by the server: $\mathcal{P}^{\mathrm{fused}}
=
\left\{(
\mathbf{c}_{\texttt{key}}^{\mathrm{fused}},
M_{\texttt{key}}^{\mathrm{fused}},
C_{\texttt{key}}^{\mathrm{fused}}) \right\}$, where $\texttt{key} = (c, (i,j))$ denotes the class-bin key, $\mathbf{c}_{\texttt{key}}^{\mathrm{fused}}$ is the encrypted fused moment
vector, and $M_{\texttt{key}}^{\mathrm{fused}}$ and $C_{\texttt{key}}^{\mathrm{fused}}$
are the associated statistics obtained during the fusion stage.

\paragraph{Moment Decryption and Parameter Recovery:}

For each \texttt{key}, $v$ decrypts the fused moment vector:
\[
\mathbf{s}_{\texttt{key}} =
\mathsf{Dec}_{\mathsf{HE}}\!\left(
\mathbf{c}_{\texttt{key}}^{\mathrm{fused}}
\right).
\]

Let
\[
\mathbf{s}_{\texttt{key}} =
\begin{bmatrix}
S_w,
S_{w\mu_x},
S_{w\mu_x^2},
S_{w\sigma_x^2},
S_{w\mu_y},
S_{w\mu_y^2},
S_{w\sigma_y^2}
\end{bmatrix}.
\]

Using the moment inversion relations derived in Lemma~\ref{lem:moment-linearity}, the Gaussian parameters of the fused detection are recovered as:

\[
\mu_x = \frac{S_{w\mu_x}}{S_w},
\qquad
\mu_y = \frac{S_{w\mu_y}}{S_w},
\]

\[
\sigma_x^2 =
\frac{S_{w\sigma_x^2}+S_{w\mu_x^2}}{S_w}-\mu_x^2,
\qquad
\sigma_y^2 =
\frac{S_{w\sigma_y^2}+S_{w\mu_y^2}}{S_w}-\mu_y^2.
\]

The corresponding standard deviations are

\[
\sigma_x=\sqrt{\max(\sigma_x^2,\epsilon)}, \qquad
\sigma_y=\sqrt{\max(\sigma_y^2,\epsilon)} .
\]
where $\epsilon>0$ is a small numerical constant used to ensure
non-negative variance and maintain numerical stability during
moment inversion.

\paragraph{Bounding Box Reconstruction:}
The recovered Gaussian parameters define a spatial extent for the
fused detection. This Gaussian representation is converted into a
bounding box $\mathbf{b} = [x_1,y_1,x_2,y_2]$, such that $x_1 = \mu_x - \lambda\sigma_x$, $y_1 = \mu_y - \lambda\sigma_y$, $x_2 = \mu_x + \lambda\sigma_x$ and $y_2 = \mu_y + \lambda\sigma_y$, 
where $\lambda$ controls the spatial coverage of the reconstructed box.

\paragraph{Class-wise Spatial Hypothesis Grouping:}
Since soft assignment distributes moment contributions across
neighboring bins, the same physical object may appear in multiple
adjacent class-bin entries. To recover a consistent set of detections, entries are grouped by object class and spatial consistency is analyzed between neighboring bins. For each class $c$, let $\mathcal{E}_c = \left\{ (\texttt{key},\mu_x,\mu_y,\sigma_x,\sigma_y,\mathbf{b},\mathbf{s}_{\texttt{key}})
\right\}$ denote the set of reconstructed bin-level entries.

\paragraph{Spatial Consistency Graph:}
For each class $c$, we construct an undirected graph given by: $\mathcal{G}_c = (\mathcal{V}_c, \mathcal{E}_c^{\mathrm{graph}})$,
where the vertex set $\mathcal{V}_c = \mathcal{E}_c$ consists of the
reconstructed bin-level entries for class $c$ (i.e., each vertex $\texttt{key} \in \mathcal{V}_c$ represents a local hypothesis $\mathcal{H}_{\texttt{key}}$ associated with a class-bin entry). An edge is introduced between two vertices $\texttt{key}_a, \texttt{key}_b \in \mathcal{V}_c$ if their corresponding bins are spatially adjacent and their associated Gaussian statistics satisfy spatial consistency constraints. Specifically, an edge $(\texttt{key}_a,\texttt{key}_b) \in \mathcal{E}_c^{\mathrm{graph}}$ is added if following condition holds:
\begin{itemize}
    \item Adjacency: the bin indices $(i,j)$ of $\texttt{key}_a$ and $\texttt{key}_b$ satisfy: $|i_a - i_b| \le 1$ and $|j_a - j_b| \le 1$.
    \item Let $\gamma_x$ and $\gamma_y$ be tunable gating constants that control the allowable spatial deviation between neighboring hypotheses. Then, the difference between Gaussian centers is bounded relative to their spatial uncertainty, as given below:
    \begin{equation}
        \label{eq:adjacency}
        \begin{split}
            |\mu_x^{(a)} - \mu_x^{(b)}| &\leq \gamma_x \min(\sigma_x^{(a)},\sigma_x^{(b)}), \\
            |\mu_y^{(a)} - \mu_y^{(b)}| &\leq \gamma_y \min(\sigma_y^{(a)},\sigma_y^{(b)}).
        \end{split}
    \end{equation}
    \item The hypotheses exhibit sufficient geometric overlap and
    statistical proximity, defined as follows:
        \begin{itemize}
            \item \textit{Geometric overlap:} The intersection-over-union (IoU)
            between the bounding boxes satisfies: $$
            \mathrm{IoU}(\mathbf{b}^{(a)}, \mathbf{b}^{(b)}) \ge \tau_{\mathrm{high}}$$
            \item \textit{Statistical proximity:} The squared Mahalanobis
            distance between the Gaussian parameters satisfies: $$
            m^2\bigl((\mu^{(a)},\Sigma^{(a)}),(\mu^{(b)},\Sigma^{(b)})\bigr)
            \le \tau_m \quad \text{and} \quad \mathrm{IoU}(\mathbf{b}^{(a)}, \mathbf{b}^{(b)}) \ge \tau_{\mathrm{min}}.$$
        \end{itemize}
        where $\tau_{\mathrm{high}}$ and $\tau_{\mathrm{min}}$ denote high and minimum overlap thresholds, respectively, and $\tau_m$ controls statistical similarity.
\end{itemize}

\begin{figure}[t]
\centering
\begin{tikzpicture}[>=latex, font=\small]

\draw (0,0) rectangle (3,3);
\draw (1,0) -- (1,3);
\draw (2,0) -- (2,3);
\draw (0,1) -- (3,1);
\draw (0,2) -- (3,2);

\node at (0.5,2.5) {$\mathcal{H}_1$};
\node at (1.5,2.5) {$\mathcal{H}_2$};
\node at (1.5,1.5) {$\mathcal{H}_3$};

\node[font=\bfseries] at (1.5,3.4) {Bin-level hypotheses};

\draw[->, thick] (3.4,1.5) -- (4.6,1.5);

\node[circle,draw] (a) at (5.2,2.2) {$\mathcal{H}_1$};
\node[circle,draw] (b) at (6.5,2.2) {$\mathcal{H}_2$};
\node[circle,draw] (c) at (5.8,1.0) {$\mathcal{H}_3$};

\draw (a)-- node[above, font=\scriptsize] {IoU $\uparrow$} (b);
\draw (a)-- node[left, font=\scriptsize] {$m^2 \downarrow$} (c);
\draw (b)-- (c); 

\node[font=\bfseries] at (5.8,3.4) {Consistency graph};

\draw[->, thick] (7.2,1.5) -- (8.4,1.5);

\node[draw,rounded corners,minimum width=2cm,minimum height=1.2cm]
at (9.7,1.5) {Merged Detection};

\node[font=\bfseries] at (9.7,3.4) {Cluster fusion};

\end{tikzpicture}

\caption{Cross-bin hypothesis merging. Neighboring bins produce Gaussian hypotheses that may correspond to the same physical object. A spatial consistency graph is constructed between hypotheses and connected components are merged together producing the final fused detection.}
\label{fig:cross-bin-merge}
\end{figure}
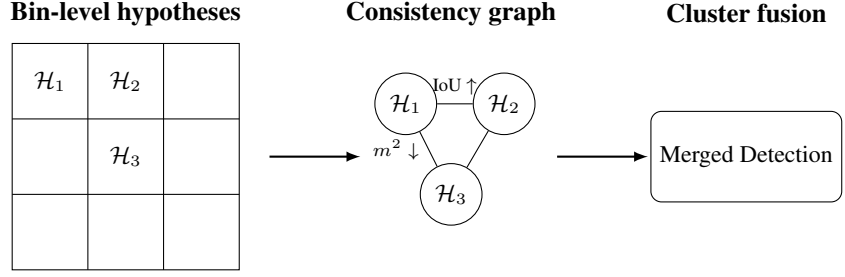

Let $\mathcal{K}_c = \{K_1, K_2, \dots\}$ denote the set of connected components of $\mathcal{G}_c$, where each $K \subseteq \mathcal{V}_c$ is a set of vertices (i.e., class-bin keys). Each component $K$ therefore represents a merged object hypothesis formed by grouping spatially-consistent local hypotheses $\{\mathcal{H}_{\texttt{key}} \mid \texttt{key} \in K\}$. Figure~\ref{fig:cross-bin-merge} depicts the construction of the spatial consistency graph and subsequent cluster fusion. Local hypotheses $\mathcal{H}_\texttt{key}$ arising from neighboring bins are linked based on geometric overlap and statistical proximity, and connected components are aggregated to produce the final detection.

\paragraph{Cluster Fusion and Confidence Recovery.}

For each connected component $K \in \mathcal{K}_c$, merged hypothesis is constructed by aggregating the moment vectors of all participating entries, hence we have: $$\mathbf{s}^{K} = \sum_{\texttt{key} \in K} \mathbf{s}_{\texttt{key}}.$$
The Gaussian parameters $(\mu_x^K,\mu_y^K,\sigma_x^K,\sigma_y^K)$ and the corresponding bounding box $\mathbf{b}^K$ are then recovered from $\mathbf{s}^K$ using the moment inversion and Gaussian-to-bounding-box mapping described previously.

\begin{figure}[t]
     \centering
     \begin{subfigure}[b]{0.49\textwidth}
         \centering
         \includegraphics[width=\textwidth]{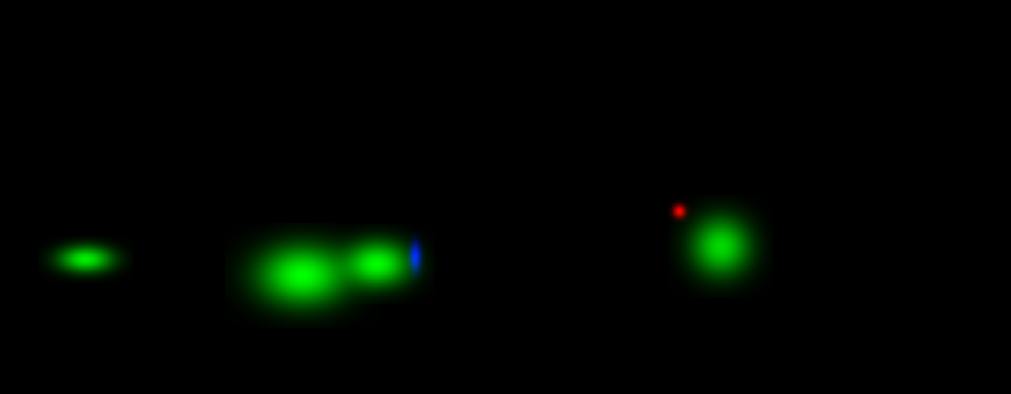}
         \caption{Gaussian splat after HE merging of RT-DETR and DETR101.}
         \label{fig-rt-detr-detr101-splat}
     \end{subfigure}
     %\hfill
     \begin{subfigure}[b]{0.49\textwidth}
         \centering
         \includegraphics[width=\textwidth]{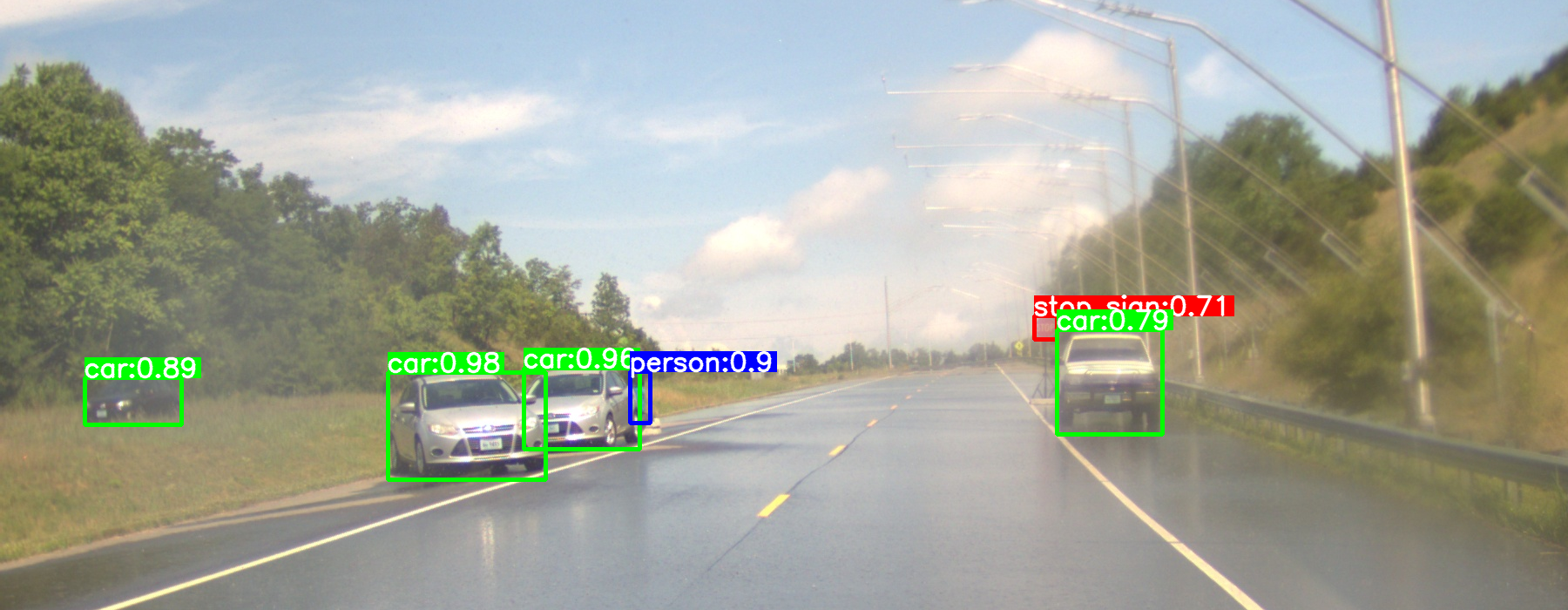}
         \caption{Bounding box obtained after cluster fusion.}
         \label{fig-rt-detr-detr101-detection}
     \end{subfigure}
    \caption{HE Fusion: RT-DETR and DETR101.}
    \label{fig-rt-detr-detr101-splat-detection}
\end{figure}

The associated confidence statistics are obtained from the fused payload by aggregating the corresponding quantities across the cluster: 
$$\hat{m}^{K} =
\sum_{\texttt{key} \in K} M_{\texttt{key}}^{\mathrm{fused}},
\qquad
\hat{c}^{K} =
\sum_{\texttt{key} \in K} C_{\texttt{key}}^{\mathrm{fused}}.$$
The fused confidence score for the hypothesis is therefore: $
\hat{p}^{K} = \frac{\hat{m}^{K}}{\hat{c}^{K}}$. The final fused detection set is given by:
$$
\mathcal{O}
=
\left\{
(\mu_x^K,\mu_y^K,\sigma_x^K,\sigma_y^K,\mathbf{b}^K,\hat{m}^{K},\hat{p}^{K})
\;\middle|\;
K \in \mathcal{K}_c,\; c \in \mathcal{C}
\right\},
$$
such that each element corresponds to a reconstructed object obtained from merging a cluster of spatially consistent hypotheses. This completes the reconstruction pipeline, mapping encrypted multi-vendor observations to a consistent set of fused object detections. Algorithm~\ref{alg:fused-payload-to-fused-detection} summarizes the complete methodology discussed in this section in pseudocode form. Figure~\ref{fig-rt-detr-detr101-splat-detection} shows the obtained Gaussian splat (Figure~\ref{fig-rt-detr-detr101-splat}) after HE merging and the respective detection obtained (Figure~\ref{fig-rt-detr-detr101-detection}) after cluster fusion.

\begin{algorithm}[!t]
\caption{Encrypted Fused Payload to Fused Detection}
\label{alg:fused-payload-to-fused-detection}
    \begin{algorithmic}[1]
        \Require 
            \parbox[t]{0.9\linewidth}{
            Server fused payload $\mathcal{P}^{\mathrm{fused}}$:\\
            \quad $\texttt{fused\_cipher\_by\_key}$: $\{\texttt{key}:(\texttt{class\_id} , \texttt{bin\_id}), \texttt{ciphertext\_bytes}\}$ \Comment{Fused encrypted  $\mathbf{v}$}\\
            \quad $\texttt{fused\_mass\_by\_key}$: $\{\texttt{key}:(\texttt{class\_id} , \texttt{bin\_id)}, \texttt{float}\}$ \Comment{Fused confidence mass}\\
            \quad $\texttt{fused\_count\_by\_key}$: $\{\texttt{key}:(\texttt{class\_id} , \texttt{bin\_id)}, \texttt{float}\}$ \Comment{Fused detection count}
        }
        \State $\texttt{bin\_by\_class} \leftarrow \emptyset$ \Comment{Group bins by class}
        \State $\mathcal{O} \leftarrow \emptyset$ \Comment{Fused detections}
        \For{each $(\texttt{key}, \texttt{ciphertext\_bytes}) \in \texttt{fused\_cipher\_by\_key}$}
            \State $(\texttt{class\_id} , \texttt{bin\_id}) \leftarrow \texttt{key}$
            \State $\mathbf{s} \leftarrow \textsf{Dec}_{\mathsf{HE}}(\texttt{ciphertext\_bytes})$ \Comment{Homomorphic decryption}
            \If{$\mathbf{s}[0] \leq \epsilon$}
                \State $\texttt{continue}$
            \EndIf
            \State ($\mu_x , \mu_y, \sigma_x, \sigma_y) \leftarrow \textbf{InvertMoment}(\mathbf{s})$
            \State $\mathbf{b} \leftarrow \textbf{GaussianToBBox}(\mu_x , \mu_y, \sigma_x, \sigma_y, \lambda)$

            \If{$\text{len}(\texttt{bin\_by\_class} [\texttt{class\_id}]) = 0$} 
                \State $\texttt{bin\_by\_class}[\texttt{class\_id}] \leftarrow [\texttt{class\_id}, \texttt{bin\_id}, \mathbf{s}[0], \mu_x,\mu_y, \sigma_x, \sigma_y, \mathbf{b}, \mathbf{s}]$ 
            \Else
                \State $\texttt{bin\_by\_class}[\texttt{class\_id}].\text{append}( [\texttt{class\_id}, \texttt{bin\_id}, \mathbf{s}[0], \mu_x,\mu_y, \sigma_x, \sigma_y, \mathbf{b}, \mathbf{s}])$
            \EndIf
        \EndFor

        \For{each $(c, \mathcal{E}_c) \in \texttt{bin\_by\_class}$}
            \State $\mathcal{O}_c \leftarrow \textbf{Merge}(\mathcal{E}_c,\texttt{fused\_mass\_by\_key},\texttt{fused\_count\_by\_key})$
            \State $\mathcal{O} \leftarrow \mathcal{O} \cup \mathcal{O}_c$
        \EndFor
        \State \textbf{Return} $\mathcal{O}$
        
    \end{algorithmic}
    \hfill\\
    \textbf{InvertMoments}$(\mathbf{s})$\textbf{:}
    \begin{algorithmic}[1]
        \State $[S_w, S_{wc_x}, S_{wc_x^2}, S_{w\sigma_x^2}, S_{wc_y}, S_{wc_y^2}, S_{w\sigma_y^2}] \leftarrow \mathbf{s}$
        \State $\mu_x \gets S_{wc_x} / S_w$, $\mu_y \gets S_{wc_y} / S_w$
        \State $v_x \gets (S_{w\sigma_x^2} + S_{wc_x^2}) / S_w - \mu_x^2$, $v_y \gets (S_{w\sigma_y^2} + S_{wc_y^2}) / S_w - \mu_y^2$
        \State $v_x \gets \max(v_x, \epsilon)$, $v_y \gets \max(v_y, \epsilon)$
        \State $\sigma_x \gets \sqrt{v_x}$, $\sigma_y \gets \sqrt{v_y}$
        \State \textbf{Return} $(\mu_x, \mu_y, \sigma_x, \sigma_y)$
    \end{algorithmic}

    \hfill\\
    \textbf{GaussianToBBox}$(\mu_x , \mu_y, \sigma_x, \sigma_y, \lambda)$\textbf{:}
    \begin{algorithmic}[1]
        \State $x_1 \leftarrow (\mu_x - \lambda \sigma_x)$, $y_1 \leftarrow (\mu_y - \lambda \sigma_y)$ 
        \State $x_2 \leftarrow (\mu_x + \lambda \sigma_x)$, $y_2 \leftarrow (\mu_y + \lambda \sigma_y)$
        \State \textbf{Return} $[x_1, y_1, x_2, y_2]$
    \end{algorithmic}

    \hfill\\
    \textbf{Merge}$(\texttt{entries}, \texttt{fused\_mass\_by\_key}, \texttt{fused\_count\_by\_key})$\textbf{:}
    \begin{algorithmic}[1]
    \Require Per-bin entries $\texttt{entries}=\{e_i\}$ for a fixed class
        \State $\mathcal{O}_{\mathrm{cls}} \leftarrow \emptyset$
    \State $n \leftarrow |\texttt{entries}|$

    \State Initialize adjacency list $\texttt{Adj}[0 \dots n-1]$

    \ForAll{pairs $(i,j)$ with $i < j$} \Comment{Ensure undirected graph}
        \If{$\texttt{entries}[i].\texttt{bin\_id}$ and $\texttt{entries}[j].\texttt{bin\_id}$ are not neighbors}
            \State continue
        \EndIf

        \State $\Delta x \leftarrow |\texttt{entries}[i].\mu_x - \texttt{entries}[j].\mu_x|$
        \State $\Delta y \leftarrow |\texttt{entries}[i].\mu_y - \texttt{entries}[j].\mu_y|$

        \State $\sigma_{x,\min} \leftarrow \min(\texttt{entries}[i].\sigma_x,\texttt{entries}[j].\sigma_x)$
        \State $\sigma_{y,\min} \leftarrow \min(\texttt{entries}[i].\sigma_y,\texttt{entries}[j].\sigma_y)$
        
        \algstore{alg-vendorpost}
        
\end{algorithmic}
    
\end{algorithm}

\begin{algorithm}
    \begin{algorithmic}[!h]
    
    \algrestore{alg-vendorpost}

        \If{$\Delta x > \gamma_x \sigma_{x,\min}$ \textbf{or} $\Delta y > \gamma_y \sigma_{y,\min}$} \Comment{Center distance gating}
            \State continue
        \EndIf
    
    \State $IoU \leftarrow \text{IoU}(\texttt{entries}[i].\texttt{bbox},\texttt{entries}[j].\texttt{bbox})$ \Comment{IoU calculation}
        \State $m^2 \leftarrow \text{Maha}^2((\mu_i,\sigma_i),(\mu_j,\sigma_j))$ \Comment{Mahalanobis distance}

        \If{$IoU \ge \tau_{\mathrm{strong}}$ \textbf{or} $(m^2 \le \tau_m \land IoU \ge \tau_{\mathrm{floor}})$} \Comment{Geometric overlap and statistical proximity}
            \State add edge $i \leftrightarrow j$ to $\texttt{Adj}$ \Comment{Add edge from $i$ to $j$}
        \EndIf
    \EndFor
    \State $\mathcal{G} \leftarrow (\texttt{Vertex},\texttt{Edge})$ constructed from $\texttt{Adj}$ \Comment{Undirected graph: \texttt{Edge} between bins (\texttt{Vertex}) if adjacent}
    \State $\mathcal{K} \leftarrow \textbf{ConnectedComponents}(\mathcal{G})$ \Comment{Perform Breadth First Search on $\mathcal{G}$}

    \ForAll{clusters $K \in \mathcal{K}$}
        \If{$|K| = 1$}
            \State $e_k \leftarrow \texttt{entries}[k]$ \Comment{$k \leftarrow$ the single element of $K$}
            \State $M_k \leftarrow \texttt{fused\_mass\_by\_key}[(e_k.\texttt{class\_id},e_k.\texttt{bin\_id})]$
            \State $A_k \leftarrow \texttt{fused\_count\_by\_key}[(e_k.\texttt{class\_id},e_k.\texttt{bin\_id})]$
            \State $\hat{p}_k \leftarrow M_k/A_k$
            \State  $ \mathcal{O}_{\mathrm{cls}}.\text{append}(e_k.\mu_x,e_k.\mu_y,e_k.\sigma_x,e_k.\sigma_y,e_k.\mathbf{b},M_k,\hat{p}_k)$
            \State continue
        \EndIf

        \State $\mathbf{s}^{K} \leftarrow \sum_{k \in K} \texttt{entries}[k].\mathbf{s}$
        \State $(\mu_x^K,\mu_{y}^K,\sigma_x^K,\sigma_y^K) \leftarrow \textbf{InvertMoments}(\mathbf{s}^{K})$
        \State $\mathbf{b}^K \leftarrow \textbf{GaussianToBBox}(\mu_x^K,\mu_y^K,\sigma_x^K,\sigma_y^K, \lambda)$

        \If{$(\sigma_x^K > \kappa_\sigma \max_{k \in K}(\sigma_{x,k})$
            \textbf{or}
            $\sigma_y^K > \kappa_\sigma \max_{k \in K}(\sigma_{y,k}))$ \textbf{or} $(\mathrm{Area}(b^K) > \kappa_A \max_{k \in K}(\mathrm{Area}(b_k)))$}
            \For{$k \in K$}
                \State $e_k \leftarrow \texttt{entries}[k]$
                \State $M_k \leftarrow \texttt{fused\_mass\_by\_key}[(e_k.\texttt{class\_id},e_k.\texttt{bin\_id})]$
                \State $A_k \leftarrow \texttt{fused\_count\_by\_key}[(e_k.\texttt{class\_id},e_k.\texttt{bin\_id})]$
                \State $\hat{p}_k \leftarrow M_k/A_k$
                \State  $ \mathcal{O}_{\mathrm{cls}}.\text{append}(e_k.\mu_x,e_k.\mu_y,e_k.\sigma_x,e_k.\sigma_y,e_k.\mathbf{b},M_k,\hat{p}_k)$
            \EndFor
            \State continue
        \EndIf

        \State $\hat{m} \leftarrow \sum_{k \in K} \texttt{fused\_mass\_by\_key}[(\texttt{entries}[k].\texttt{class\_id}, \texttt{entries}[k].\texttt{bin\_id})]$ \Comment{Confidence mass}
        \State $\hat{c} \leftarrow \sum_{k \in K} \texttt{fused\_count\_by\_key}[(\texttt{entries}[k].\texttt{class\_id}, \texttt{entries}[k].\texttt{bin\_id})]$ \Comment{Aggregation weight}
        \State $\hat{p} \leftarrow \frac{\hat{m}}{\hat{c}}$ \Comment{Fused confidence}
        \State $\mathcal{O}_{\mathrm{cls}}.\text{append}(\mu_x^K,\mu_y^K,\sigma_x^K,\sigma_y^K, \mathbf{b}^K, \hat{m}, \hat{p})$ \Comment{Append the merged object}
    \EndFor
    
    \end{algorithmic}
\end{algorithm}

\subsection{Complexity Analysis}
\label{sec:complexity-analysis}
\subsubsection{Vendor Payload Construction}
\label{sec:vendor-payload-complexity}
Let $N$ denote the number of detections produced by a vendor and let $B_v$ denote the number of occupied class-bin keys for that vendor i.e.,$\texttt{key} = (c,(i,j))$ after spatial binning.
\begin{enumerate}
    \item \textbf{Per-Detection Processing:} For each detection, \textsf{Sarus} performs (i) Gaussian parameter computation,
    (ii) moment vector construction, and (iii) soft assignment to at most four neighboring bins. Each operation takes constant time, yielding: $$T_{\mathrm{preprocess}} = O(N).$$
    \item \textbf{Per-Key Aggregation:} Each detection contributes to at most four keys, and the corresponding moment vectors are accumulated into per-key aggregates. Since the number of updates per detection is bounded by a constant, the total cost is:$$T_{\mathrm{aggregation}} = O(N).$$
    \item \textbf{Payload Creation:} The payload is constructed by iterating over all occupied keys, resulting in: $$T_{\mathrm{payload}} = O(B_v),$$
    where typically $B_v \ll N$ due to spatial aggregation.
\end{enumerate}

Combining the above steps, the total time complexity is:
\begin{equation}
\label{eq:complexity-vendor-payload}
T_\mathrm{vendor} = O(N + B_v) = O(N).
\end{equation}
This linear complexity is achieved through local aggregation, which avoids per-detection payload expansion.

\subsubsection{Encrypted Multi-Vendor Fusion}

Let $V = |\mathcal{V}|$ denote the number of participating vendors and let $B$ denotes the union of keys across all vendors.
\begin{enumerate}
    \item \textbf{Key Collection:}
    The fusion server first computes the union of keys contributed by all vendors i.e., $\mathcal{K} = \bigcup_{v \in \mathcal{V}} \mathrm{keys}(\mathcal{P}_v)$. If each vendor contributes at most $B$ occupied keys, then collecting
    the union requires:$$T_{\mathrm{keys}} = O(BV).$$

    \item \textbf{Per-Key Fusion:}
    For each key $\texttt{key} \in \mathcal{K}$, the server scans the vendor payloads and combines the available entries. Ignoring the cost of encryption-specific operations, each per-key update is constant time: one addition for the fused moment accumulator and constant-time updates for the associated plaintext statistics. Since there are at most $B$ keys and at most $V$ vendors contributing to each key, the total cost is:$$
    T_{\mathrm{fusion}} = O(BV).$$

    \item \textbf{Fused Payload Construction:}
    After aggregation, the server stores one fused entry per occupied key, yielding: $$T_{\mathrm{output}} = O(B).$$
\end{enumerate}

Combining the above steps, the total time complexity of encrypted multi-vendor fusion by the server is:
\begin{equation}
\label{eq:complexity-server-fusion}
T_\mathrm{fusion} = O(BV + B) = O(BV).
\end{equation}
$O(BV)$ result comes from the fact that \textsf{Sarus} fuses aggregated per-key payloads, not individual detections. Without vendor-side aggregation, the server would instead process per-detection contributions, leading to a larger dependence on the number of detections. Hence, the server-side fusion complexity scales linearly with both the number of occupied class-bin keys and the number of participating multi-vendors.

\subsubsection{Fused Payload to Detection Reconstruction}

Let $B$ denote the number of occupied class-bin keys in the fused payload, and let $B_c$ denote the number of reconstructed bin-level entries for class $c$, such that:$$\sum_{c \in \mathcal{C}} B_c = B.$$
\begin{enumerate}
    \item \textbf{Per-Key Reconstruction:}
    For each fused $\texttt{key}=(c,(i,j))$, \textsf{Sarus} decrypts the fused moment vector, inverts the moments to recover $(\mu_x,\mu_y,\sigma_x,\sigma_y)$, and reconstructs the corresponding bounding box. Ignoring the cost of decryption, each of these operations takes constant time. Therefore, reconstructing all bin-level entries requires: $$T_{\mathrm{reconstruct}} = O(B).$$

    \item \textbf{Class-wise Grouping:}
    The reconstructed entries are grouped by class before graph-based merging. This requires a single pass over the fused entries, giving: $$T_{\mathrm{group}} = O(B).$$

    \item \textbf{Spatial Consistency Graph Construction:}
    For each class $c$, \textsf{Sarus} constructs a spatial consistency graph over the $B_c$ reconstructed entries. In the worst case, all pairs of entries for that class are examined to determine whether an edge should be added. Since each pairwise consistency check (bin adjacency, center gating, IoU, and Mahalanobis distance) takes constant time, the graph construction cost is: $$T_{\mathrm{graph}}^{(c)} = O(B_c^2).$$
    Summing over all classes yields
    $$T_{\mathrm{graph}} = O\!\left(\sum_{c \in \mathcal{C}} B_c^2\right),$$
    which, is upper bounded by: $$T_{\mathrm{graph}} = O(B^2).$$

    \item \textbf{Connected Components:}
    For each class-specific graph, connected components are computed using breadth-first search (BFS). BFS runs in time linear in the number of vertices and edges: $$T_{\mathrm{cc}}^{(c)} = O(B_c + E_c),$$ where $E_c$ is the number of graph edges for class $c$. Since
    $E_c = \mathcal{O}(B_c^2)$ in the worst case, this becomes: $$T_{\mathrm{cc}}^{(c)} = O(B_c^2).$$
    Therefore, $$T_{\mathrm{cc}} = O\!\left(\sum_{c \in \mathcal{C}} B_c^2\right) = O(B^2).$$

    \item \textbf{Cluster Fusion:} Each connected component $K$ is fused by summing its moment vectors, recovering Gaussian parameters, and computing the associated confidence statistics. Since each reconstructed entry belongs to exactly one connected component, the total work across all clusters is linear in the number of entries: $$T_{\mathrm{cluster}} = O(B).$$
\end{enumerate}

Combining the above steps, the total time complexity of fused payload to detection reconstruction is
\begin{equation}
\label{eq:complexity-reconstruction}
    T_\mathrm{post} = O\!\left(B + \sum_{c \in \mathcal{C}} B_c^2\right),
\end{equation}
which, in the worst case, simplifies to: $T_\mathrm{post} = O(B^2)$. Thus, reconstruction is dominated by class-wise graph construction and connected-component analysis, while the remaining stages are linear in the number of occupied fused keys. In practice, the effective cost is often substantially lower because spatial adjacency constraints limit the number of candidate pairs that can form graph edges.
\paragraph{Summary of Time Complexity:}
\begin{enumerate}
    \item \textbf{Vendor payload construction: } $T_\mathrm{vendor} = O(N + B_v)$.
    \item \textbf{Server encrypted fusion: } $T_\mathrm{fusion} = O(BV)$.
    \item \textbf{Post-fusion reconstruction: } $T_\mathrm{post} = O(B + \sum_{c \in \mathcal{C}} B_c^2)$.
\end{enumerate}
where, $N =$ detections per vendor, $V = $ vendors, $B_v = $ bins produced by one vendor, $B = $ fused bins and $B_c = $ bins per class.

\section{Experiments}
\label{sec:experiments}

The dataset was collected using a real-world autonomous test vehicle equipped with a drive-by-wire system provided by Dataspeed Inc.\footnote{\url{https://www.dataspeedinc.com/}}. The vehicle is instrumented with a multi-modal sensing suite, including a LiDAR sensor, an image sensor (Lucid Triton camera\footnote{\url{https://thinklucid.com/product/triton-16-mp-imx273/}}), and a radar sensor. In this work, only the image stream is utilized for evaluating the \textsf{Sarus} framework. Sensor data acquisition is performed using the Robot Operating System (ROS) middleware within the perception architecture described in~\cite{vassilev2026assessment}. The experiments were conducted at the Virginia Tech Transportation Institute (VTTI) autonomous vehicle testing track~\cite{vtti}, under controlled city-driving conditions.

The scene is designed to reflect realistic perception challenges. A stop sign is placed along the driving path of the ego vehicle and is partially occluded by a parked truck. Additional dynamic and static objects are present in the environment, including multiple vehicles and a pedestrian partially occluded by vegetation. To further emulate adverse environmental conditions, artificial rain is introduced during data collection.

To evaluate robustness under varying motion dynamics, the dataset is collected at three different vehicle speeds: $25$ mph ($142$ frames), $35$ mph ($100$ frames), and $55$ mph ($66$ frames), a total of $308$ frames. Increasing vehicle speed affects temporal sampling, motion blur, and relative object motion, while artificial rain introduces light scattering and attenuation effects. The combination of these factors produces realistic perception distortions that challenge detection consistency across vendors, enabling a rigorous evaluation of fusion stability under adverse and dynamic conditions.

This setup enables evaluation of perception fusion under realistic occlusion, multi-object interaction, adverse weather conditions and under varying speeds.

\paragraph{Multi-Vendor Perception Setup.}

To emulate a collaborative multi-vendor perception environment, each frame is processed using five heterogeneous object detection models: YOLOv8, YOLOv9, DETR-50, DETR-101, and RT-DETR. Each model is treated as an independent vendor providing its own set of detection outputs.

To capture varying levels of vendor participation, we evaluate $26$ non-empty subsets of these models, corresponding to combinations of $2$, $3$, $4$, and $5$ vendors. Each subset represents a distinct collaborative scenario with differing degrees of redundancy and diversity in perception. For each frame and each vendor subset, fusion is performed independently, resulting in a total of: $$308 \times 26 = 8008$$
fusion instances. This exhaustive enumeration enables a comprehensive evaluation of the proposed framework across a wide range of multi-vendor configurations. This design allows us to systematically analyze how fusion performance varies with the number and diversity of participating vendors.

\subsection{Equivalence of Homomorphic and Plaintext Fusion}

The primary objective of the experiment is to validate the \textit{functional and numerical correctness} of the proposed homomorphic fusion pipeline. Specifically, the aim is to verify that fusion performed entirely in the encrypted domain reproduces the same outputs as plaintext fusion, up to negligible numerical error induced by approximate homomorphic encryption. Such a correctness validation is a necessary prerequisite before evaluating downstream perception performance on labeled benchmarks, as any accuracy gains or losses would be meaningless without first demonstrating that encryption itself does not alter fusion behavior. 

\paragraph{Fusion Modes.}

For each frame and each vendor subset, fusion is executed under two computational settings:

\begin{enumerate}
    \item \textbf{Plaintext fusion:} the fusion pipeline operates directly on unencrypted detection outputs, serving as the reference baseline

    \item \textbf{Homomorphic fusion:} the identical fusion pipeline is executed over CKKS-encrypted representations of the detection payloads, where all aggregation operations are performed in the encrypted domain without access to plaintext data.
\end{enumerate}

In both settings, the underlying fusion logic, including moment aggregation, spatial binning, and hypothesis merging, remains unchanged. This ensures that any observed differences in output or runtime are solely attributable to the use of homomorphic encryption.

Table~\ref{tab:he-plain-equivalence} quantitatively compares the outputs of homomorphic fusion and plaintext fusion across different driving regimes. The results demonstrate near-perfect agreement between the two pipelines.

Across all speed settings, the intersection-over-union (IoU) between
bounding boxes produced by homomorphic and plaintext fusion exceeds
$0.99997$ on average, with the $95^{\text{th}}$ percentile and maximum values reaching $1.000$. This indicates that the spatial extent of the reconstructed detections is effectively identical in both settings.

The absolute deviations in bounding box centers and sizes remain within sub-pixel to pixel-level precision, with $\max|\Delta c| \leq 0.5$ pixels and $\max|\Delta s| \leq 2.0$ pixels across all experiments. Similarly, the differences in the recovered Gaussian standard deviations are negligible, with $\max|\Delta\sigma_x|$ and $\max|\Delta\sigma_y|$ remaining below $0.05$ pixels.

These small discrepancies arise from the approximate nature of CKKS-based homomorphic encryption, which introduces bounded numerical error during arithmetic operations. Importantly, the magnitude of these errors is insufficient to affect the geometric interpretation or downstream decision-making of the fused detections.

Overall, these results confirm that the proposed homomorphic fusion pipeline preserves the correctness of plaintext fusion with high numerical fidelity, validating the use of encrypted computation for privacy-preserving multi-vendor perception.

\begin{table}[t]
    %\small
    \centering
    \caption{Equivalence between homomorphic and plaintext fusion across driving regimes.}
    \label{tab:he-plain-equivalence}
    \begin{tabular}{|l|c|c|c|c|c|c|c|c|}
        \hline
        \textbf{Speed} & \textbf{N} &
        \multicolumn{3}{|c|}{\textbf{IoU(HE,Plain)}} &
        \multicolumn{2}{|c|}{\textbf{BBox $\Delta$ (px)}} &
        \multicolumn{2}{|c|}{\textbf{$\Delta\sigma$ (px)}} \\
        \cline{3-9}
         &  & mean & p95 & max & $\max|\Delta c|$ & $\max|\Delta s|$ & $\max|\Delta\sigma_x|$ & $\max|\Delta\sigma_y|$ \\
        \hline
        $25$ mph & $3692$ $(=142 \times 26)$ & $0.999989$ & $1.000$ & $1.000$ & $0.5$ & $1.0$ & $0.048$ & $0.022$ \\
        $35$ mph & $2600$ $(=100 \times 26)$ & $0.999978$ & $1.000$ & $1.000$ & $0.5$ & $1.0$ & $0.016$ & $0.005$ \\
        $55$ mph & $1716$ $(=66 \times 26)$ & $0.999991$ & $1.000$ & $1.000$ & $0.0$ & $2.0$ & $0.016$ & $0.001$ \\
        \hline
    \end{tabular}
\end{table}

\subsection{Computational Performance Evaluation}
\label{sec:exp:performance}
To evaluate the computational behavior of \textsf{Sarus}, we conduct a controlled scaling study that varies both the number of detections per scene and the number of participating vendors. Specifically, we consider detection counts: $$N \in \{1, 5, 10, 15, 20, 25\},$$
which correspond to increasing scene complexity, ranging from sparse to dense object configurations. For each setting, we simulate a collaborative perception environment with: $$V \in \{2, 3, 4, 5\}$$
independent vendors, where each vendor contributes its own detection outputs.

This setup allows us to systematically analyze how the computational cost of the pipeline scales with (i) the number of objects in the scene and (ii) the number of participating vendors. All measurements are reported separately for the three major stages of the pipeline:
(i) vendor-side payload construction,
(ii) encrypted multi-vendor fusion at the server, and
(iii) fused detection reconstruction.

Unless otherwise stated, cryptographic primitives (encryption and decryption) are excluded when analyzing algorithmic scaling behavior, in order to isolate the structural complexity of the proposed method.

\subsubsection{Vendor-Side Payload Construction}
\label{sec:exp:vendor}

\begin{figure}
     \centering
     \begin{subfigure}[b]{0.48\textwidth}
         \centering
         \includegraphics[width=\textwidth]{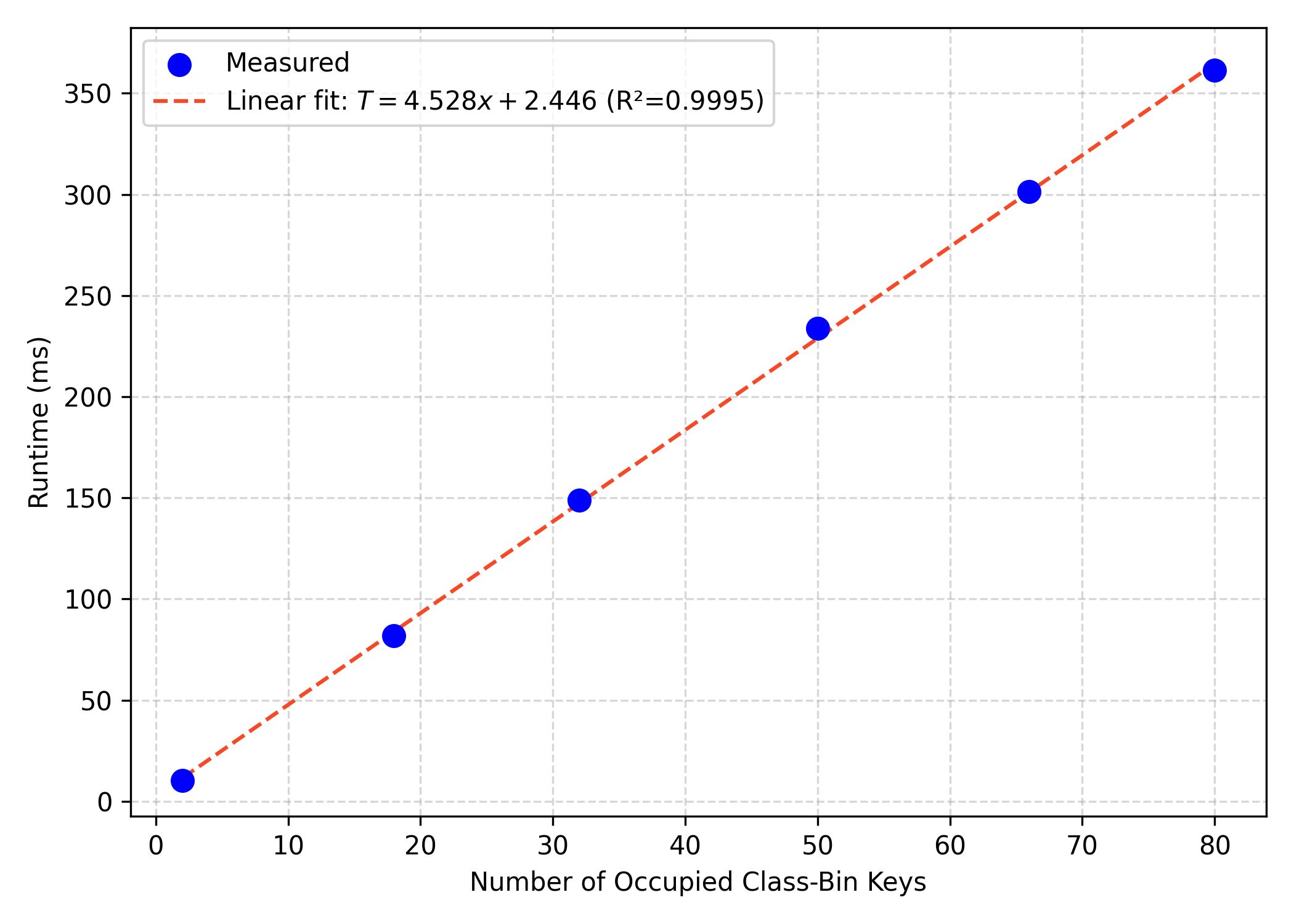}
         \caption{Total runtime as a function of occupied class-bin keys, showing a strong linear relationship ($R^2 = 0.9995$).}
         \label{fig-exp-he-payload}
     \end{subfigure}
     \hfill
     \begin{subfigure}[b]{0.48\textwidth}
         \centering
         \includegraphics[width=\textwidth]{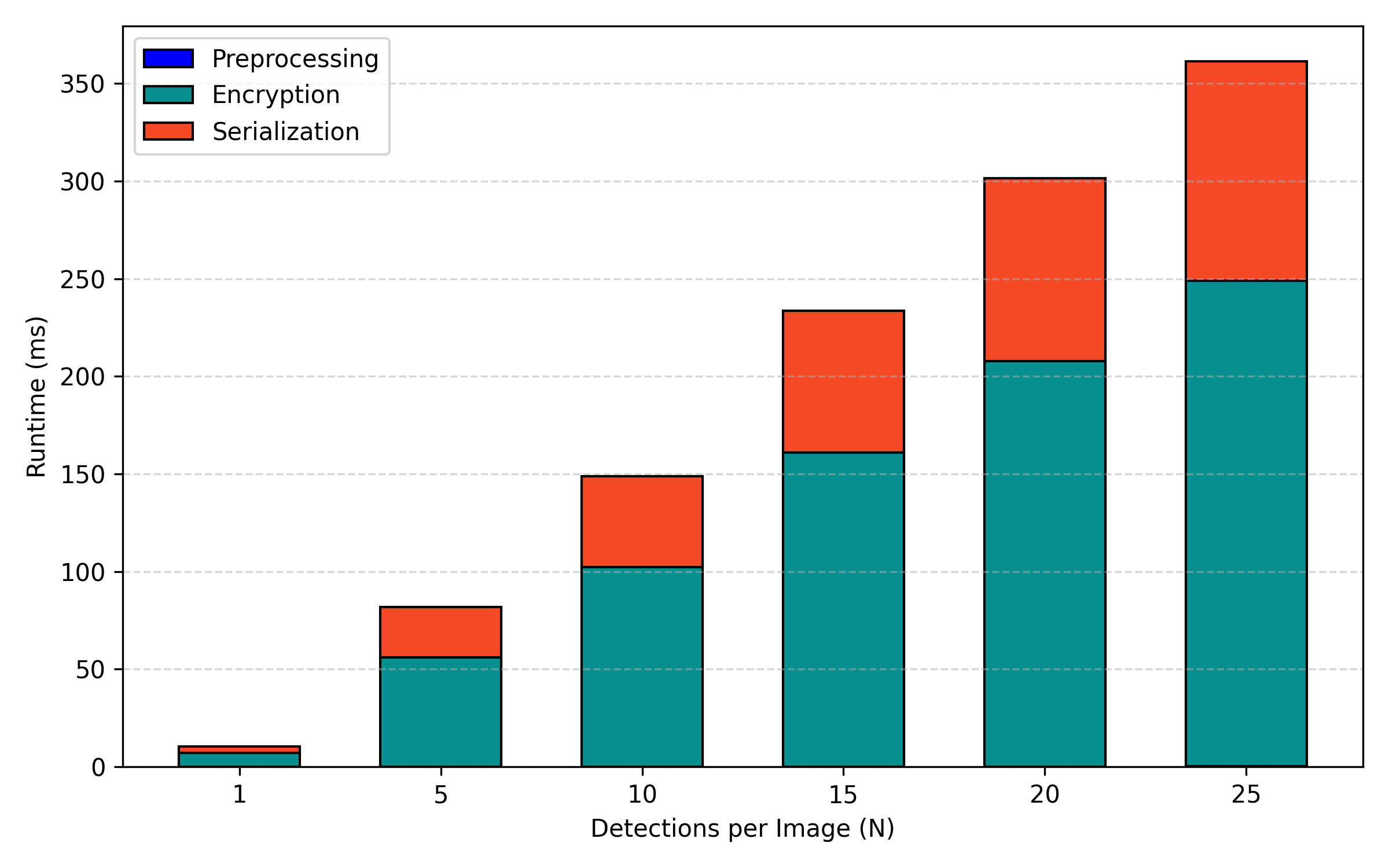}
         \caption{Stage-wise runtime breakdown highlighting encryption as the dominant cost.}
         \label{fig-exp-staged-he-payload}
     \end{subfigure}
     \hfill
     
     \begin{subfigure}[b]{0.48\textwidth}
         \centering
         \includegraphics[width=\textwidth]{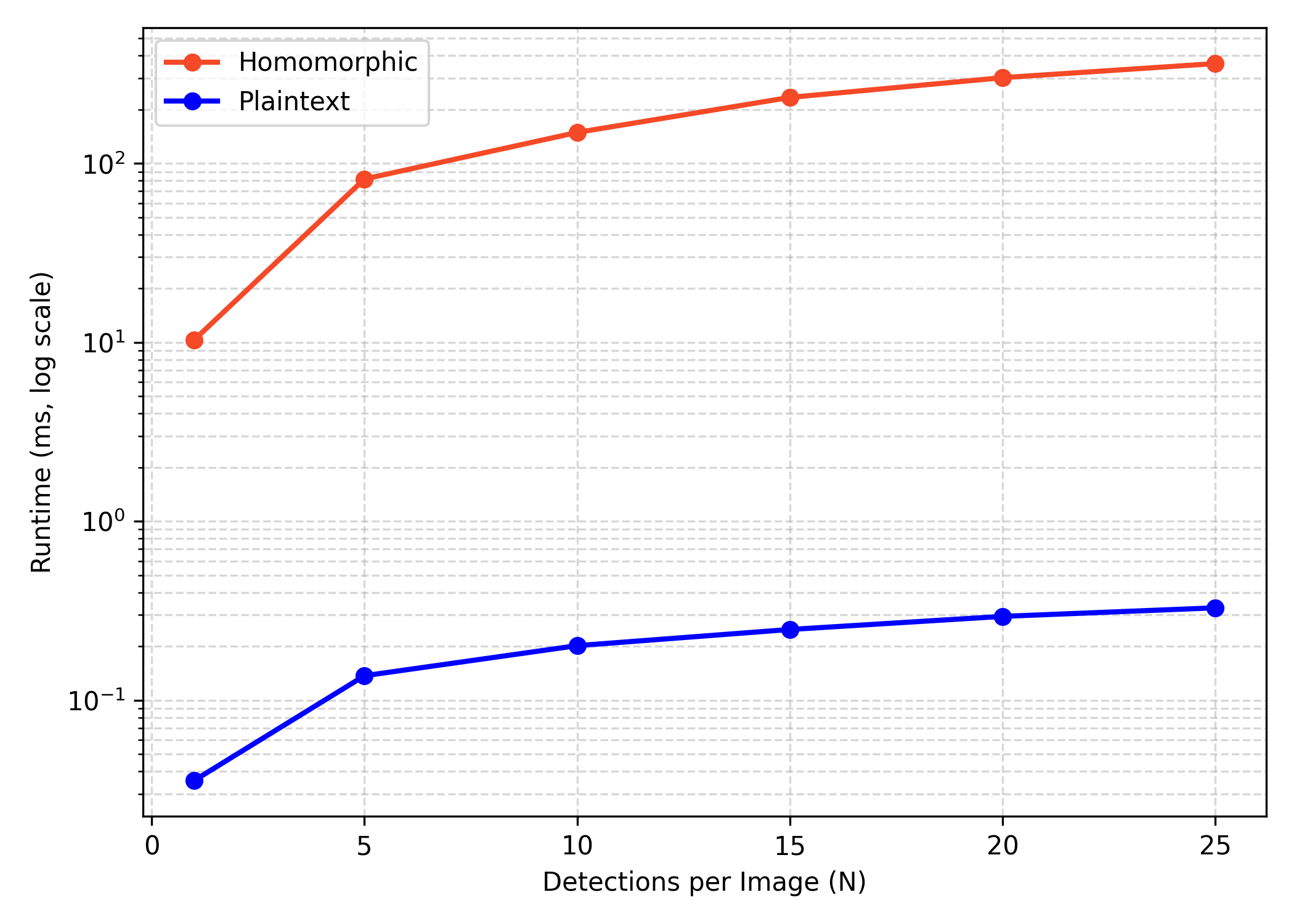}
         \caption{Comparison between homomorphic and plaintext payload construction (log scale).}
         \label{fig-exp-he-vs-plaintext}
     \end{subfigure}
     \hfill
     \begin{subfigure}[b]{0.48\textwidth}
         \centering
         \includegraphics[width=\textwidth]{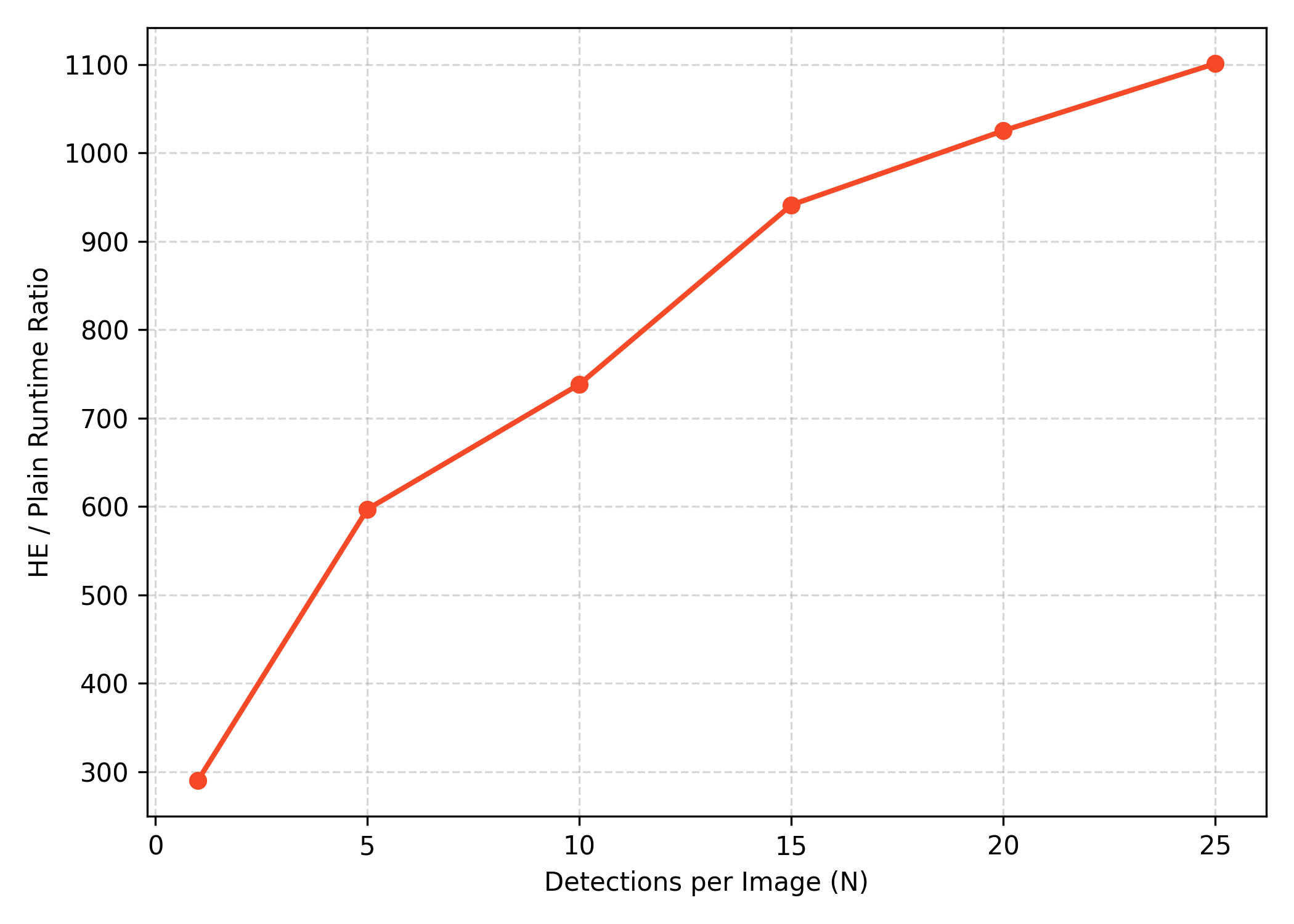}
         \caption{Relative overhead of homomorphic encryption compared to plaintext.}
         \label{fig-exp-he-payload-relative}
     \end{subfigure}
        \caption{Vendor-side payload construction performance.}
        \label{fig-exp-vendor-payload-graphs}
\end{figure}

\paragraph{Runtime vs. Occupied Bins.}
Figure~\ref{fig-exp-he-payload} shows the relationship between payload construction time and the number of occupied class-bin keys $B$. The results exhibit an almost perfectly linear trend, with a fitted model $T = 4.528\cdot B + 2.446$ and $R^2 = 0.9995$.

This provides strong empirical validation that vendor-side runtime scales as $O(B)$, confirming that the dominant cost arises from per-bin operations such as encryption and serialization, rather than from the raw number of detections.

\paragraph{Stage-wise Runtime Breakdown.}
Figure~\ref{fig-exp-staged-he-payload} decomposes runtime into preprocessing, encryption, and serialization. Encryption clearly dominates across all detection counts, accounting for the majority of total runtime (e.g., $\approx 250$ ms out of $\approx 360$ ms at $N=25$). Preprocessing remains negligible, while serialization contributes a moderate but consistent overhead.

Importantly, both encryption and serialization scale proportionally with $B$, reinforcing that cryptographic operations are the primary performance bottleneck.

\paragraph{Homomorphic vs. Plaintext Runtime.}
Figure~\ref{fig-exp-he-vs-plaintext} compares homomorphic and plaintext payload construction on a logarithmic scale. While both exhibit increasing trends with $N$, homomorphic execution is several orders of magnitude slower due to CKKS encryption overhead. However, both curves remain approximately linear, indicating that homomorphic processing preserves the same asymptotic behavior as plaintext execution.

\paragraph{Relative Overhead of Homomorphic Encryption.}
Figure~\ref{fig-exp-he-payload-relative} shows the ratio of homomorphic to plaintext runtime. The overhead ranges from approximately $300\times$ at low detection counts to over $1000\times$ at higher densities. This increase is driven by the growth in occupied bins, which amplifies the number of expensive encryption operations. Despite this overhead, the linear scaling behavior ensures predictable performance, making the system amenable to optimization through batching, parameter tuning, or hardware acceleration.

\paragraph{Discussion.}
Overall, the vendor-side evaluation confirms that payload construction scales linearly with the number of occupied bins rather than the number of detections. This distinction is critical, as spatial binning effectively compresses multiple detections into fewer encrypted representations, reducing the number of cryptographic operations.

While homomorphic encryption introduces significant constant-factor overhead, the preservation of linear scaling ensures that the system remains predictable and scalable. These results validate the theoretical complexity of $O(N) + O(B)$ and highlight spatial aggregation as a key mechanism for controlling computational cost.

\paragraph{Network Overhead:}
The use of homomorphic encryption introduces a significant increase in payload size compared to plaintext representations due to ciphertext expansion and encoding overhead. In our experiments, the encrypted payload size grows from approximately $0.65$ MB at $N=1$ to over $26$ MB at $N=25$, whereas the corresponding plaintext payload remains below $10$ KB. 

This results in a ciphertext expansion factor of approximately $3\times10^3$, which stabilizes as the number of detections increases. Despite this large constant-factor overhead, the payload size grows approximately linearly with the number of occupied bins, ensuring predictable network cost as a function of scene complexity. Furthermore, the bin-wise aggregation strategy limits the number of transmitted ciphertexts, preventing excessive growth in communication overhead even in dense detection scenarios. This predictable scaling behavior is critical for deployment, as it enables accurate estimation of bandwidth requirements despite the inherent overhead of homomorphic encryption.

\subsubsection{Server-Side Encrypted Fusion}
\label{sec:exp:server}

\begin{figure}
     \centering
     \begin{subfigure}[b]{0.48\textwidth}
         \centering
         \includegraphics[width=\textwidth]{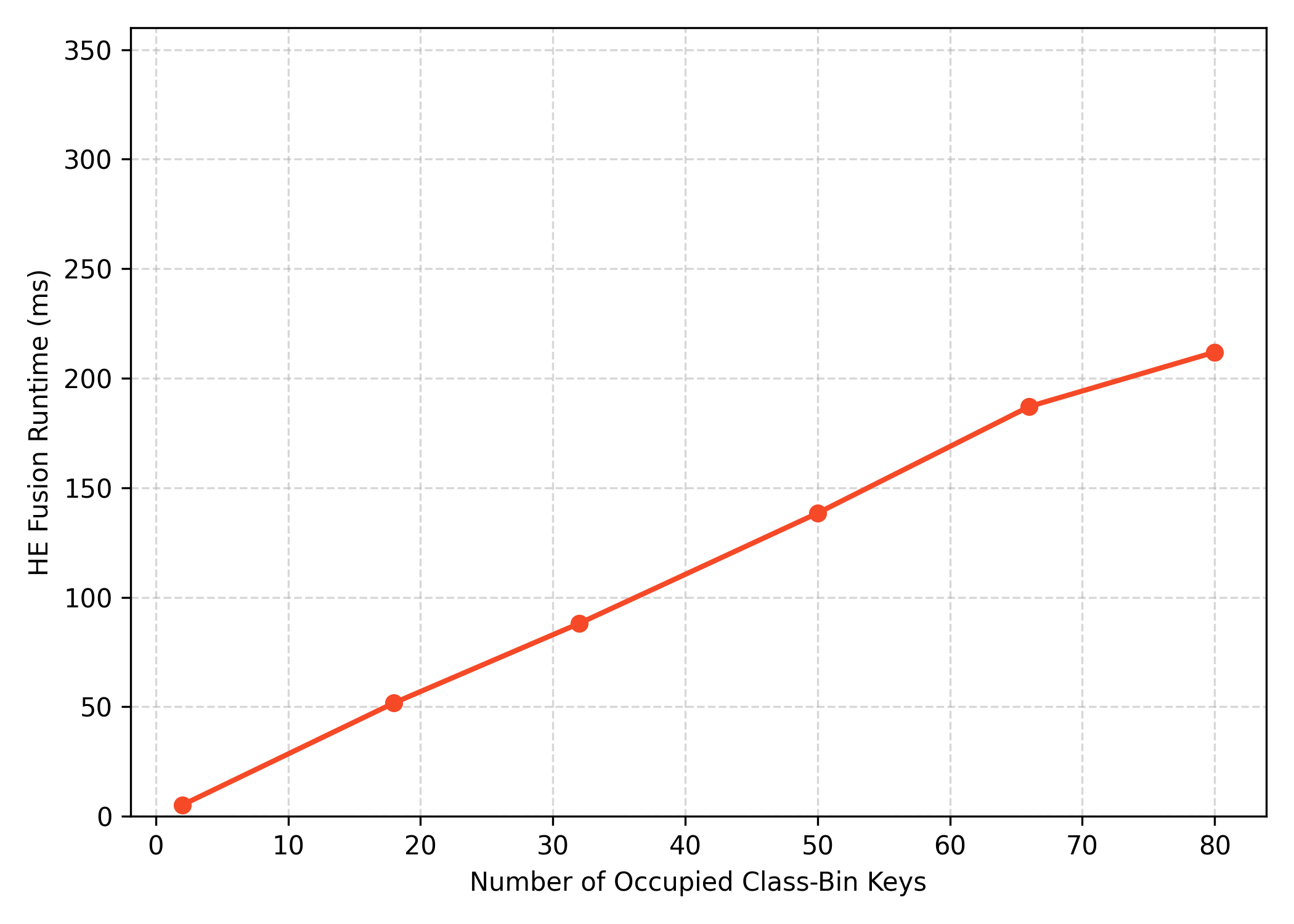}
         \caption{HE fusion runtime as a function of occupied class-bin keys for $V=2$.}
         \label{fig-exp-he-fusion-v2}
     \end{subfigure}
     \hfill
     \begin{subfigure}[b]{0.48\textwidth}
         \centering
         \includegraphics[width=\textwidth]{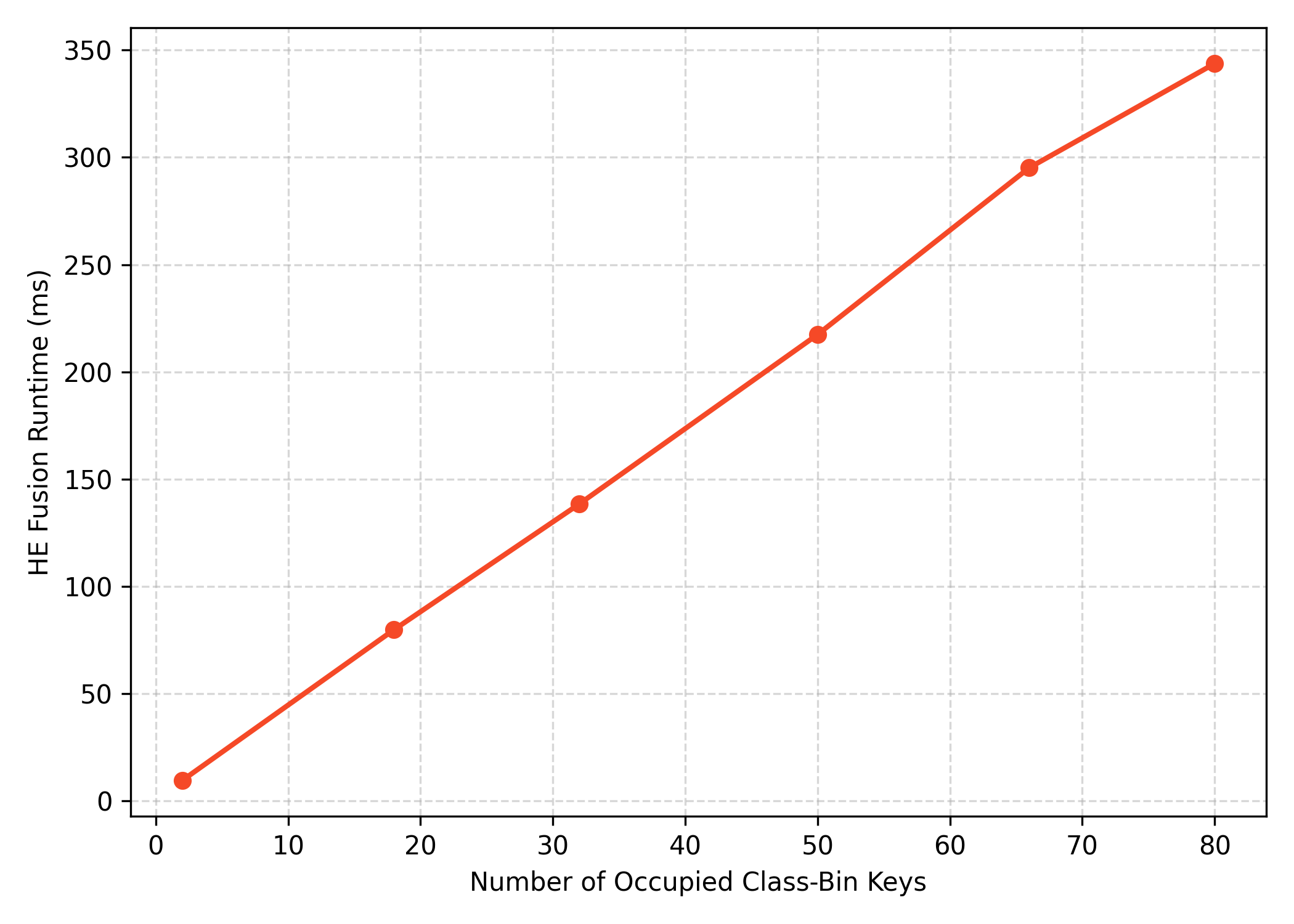}
         \caption{HE fusion runtime as a function of occupied class-bin keys for $V=5$.}
         \label{fig-exp-he-fusion-v5}
     \end{subfigure}
     \hfill
     
     \begin{subfigure}[b]{0.48\textwidth}
         \centering
         \includegraphics[width=\textwidth]{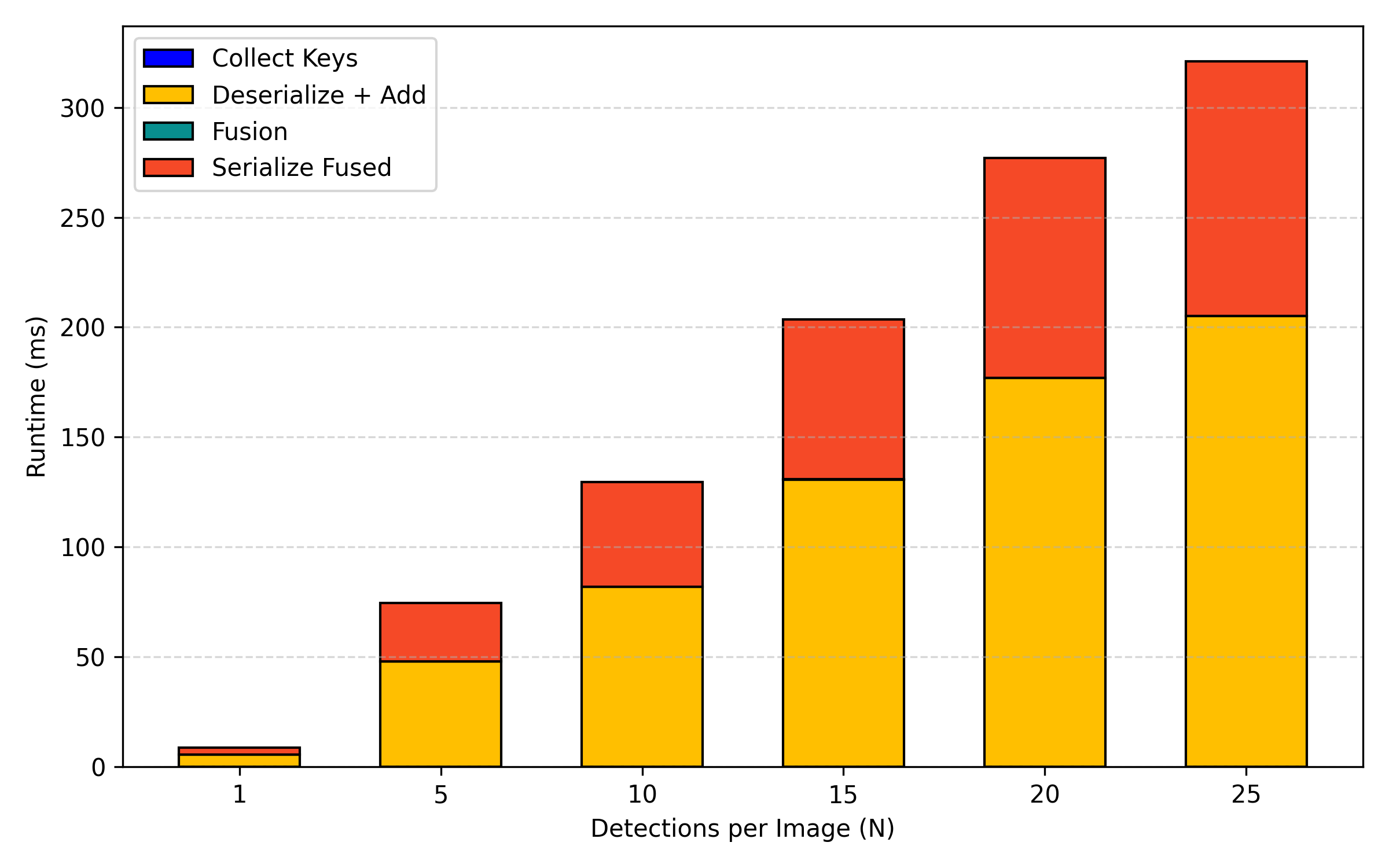}
         \caption{Stage-wise breakdown highlighting ciphertext deserialization and serialization as the dominant costs.}
         \label{fig-exp-staged-he-fusion}
     \end{subfigure}
     \hfill
     \begin{subfigure}[b]{0.48\textwidth}
         \centering
         \includegraphics[width=\textwidth]{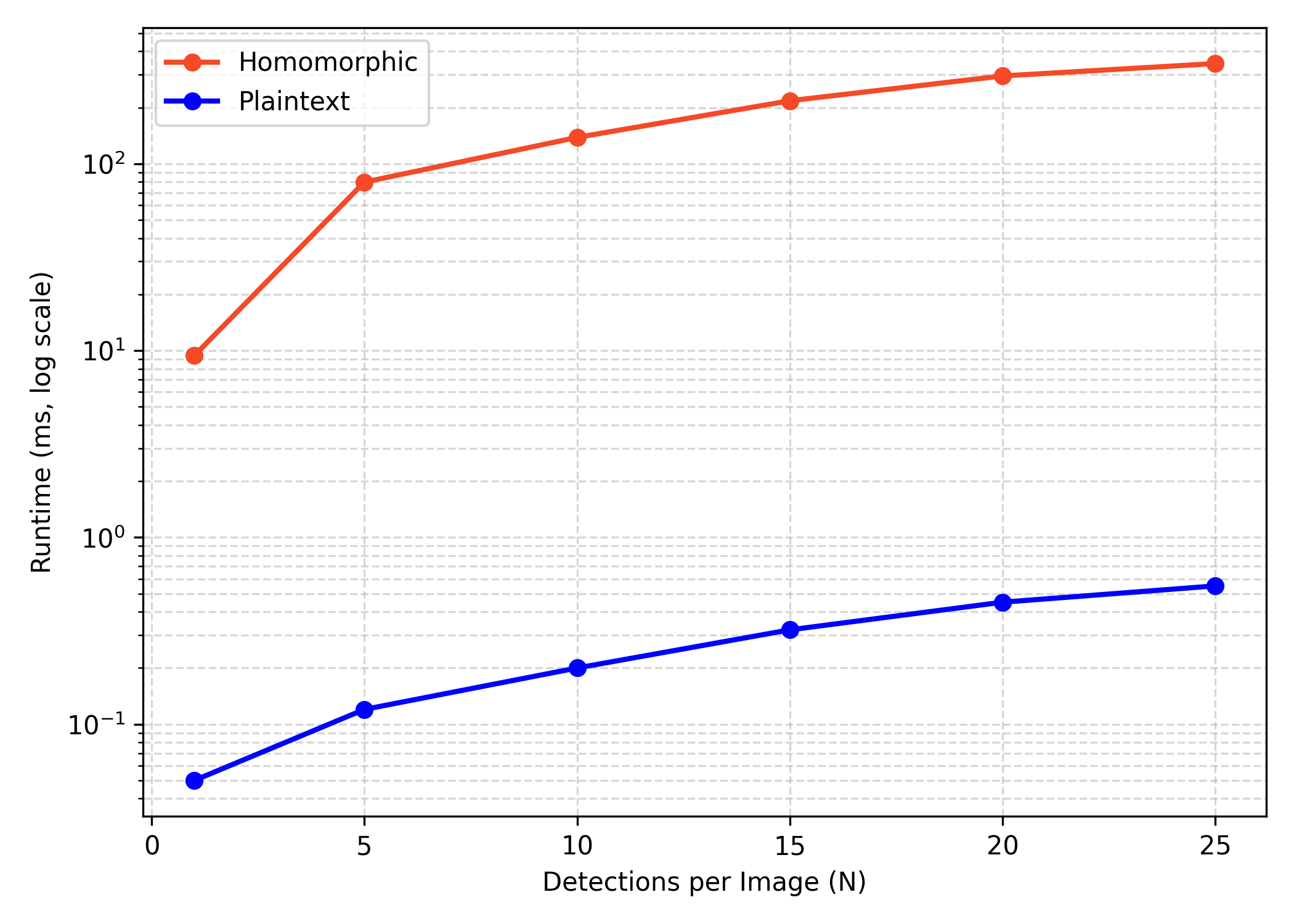}
         \caption{Comparison: homomorphic vs plaintext fusion (log scale), showing constant-factor overhead of encrypted computation.}
         \label{fig-exp-he-fusion-vs-plaintext}
     \end{subfigure}
        \caption{Server-side fusion performance.}
        \label{fig-exp-server-fusion-graphs}
\end{figure}

Figures~\ref{fig-exp-he-fusion-v2} and~\ref{fig-exp-he-fusion-v5} show that server-side HE fusion runtime scales linearly with the number of occupied class-bin keys $B$. Comparing $V=2$ and $V=5$, the slope increases with the number of vendors, confirming that the fusion cost grows proportionally with both $B$ and $V$. 

Figure~\ref{fig-exp-staged-he-fusion} provides a stage-wise breakdown, showing that the dominant cost arises from ciphertext deserialization and accumulation, followed by serialization of the fused result. In contrast, key collection and fusion arithmetic contribute negligibly to the overall runtime. 

Figure~\ref{fig-exp-he-fusion-vs-plaintext} compares homomorphic and plaintext fusion, demonstrating that while homomorphic processing introduces a significant constant-factor overhead, both exhibit similar growth trends with increasing problem size.

Empirically, the server-side fusion runtime follows a highly linear trend with respect to the number of occupied bins, with fitted slopes of approximately $2.70$ ms/bin for $V=2$ and $4.35$ ms/bin for $V=5$ ($R^2=0.9971$ and $0.9990$, respectively), confirming the expected growth with both $B$ and the number of participating vendors $V$.

\subsubsection{Fused Detection Reconstruction}
\label{sec:exp:post}
\begin{figure}
     \centering
     \begin{subfigure}[b]{0.48\textwidth}
         \centering
         \includegraphics[width=\textwidth]{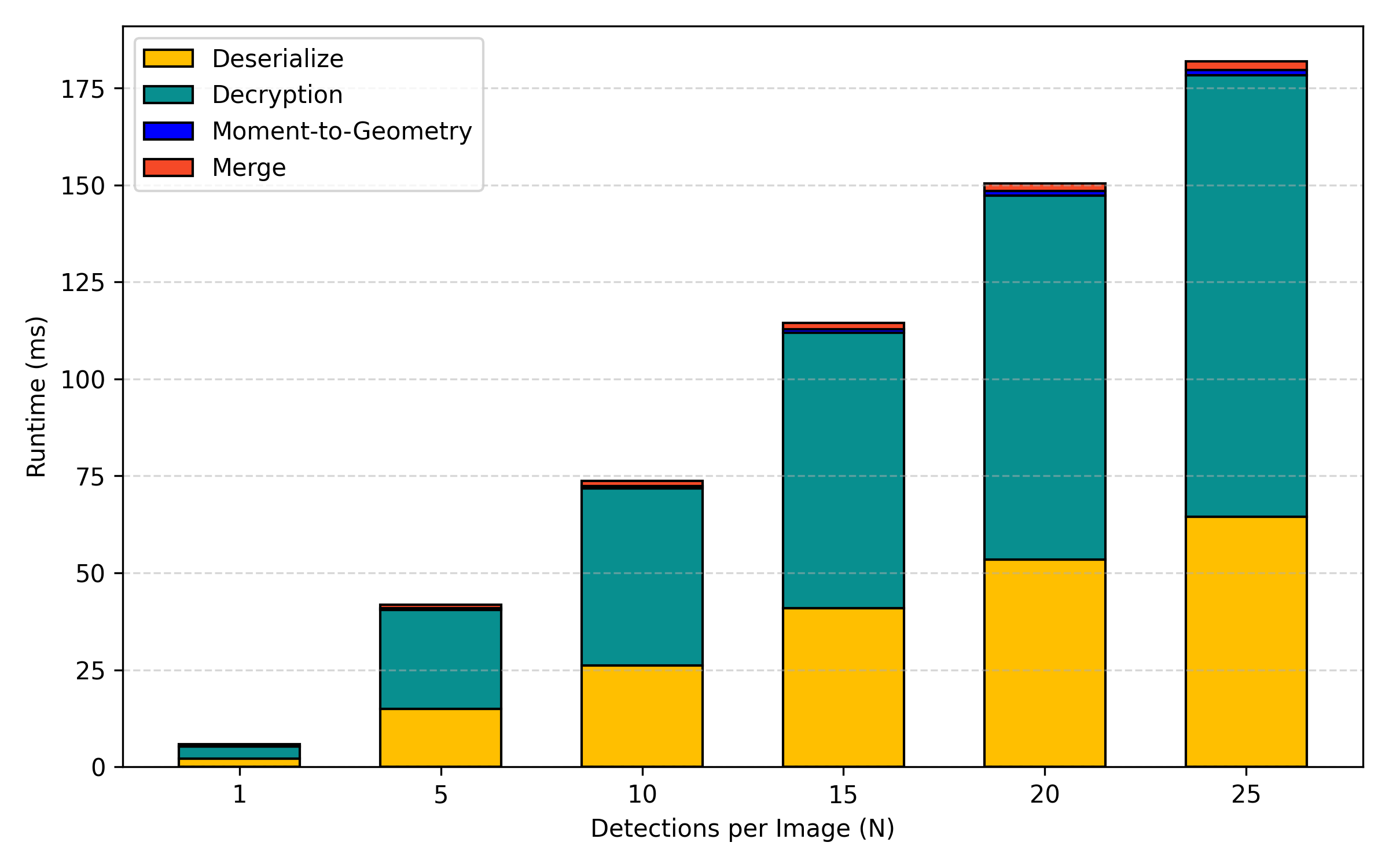}
         \caption{Stage-wise breakdown of vendor-side postprocessing under homomorphic encryption. Decryption dominates runtime, while moment-to-geometry transformation and merging incur negligible cost.}
         \label{fig-exp-vendor-post-stacked}
     \end{subfigure}
     \hfill
     \begin{subfigure}[b]{0.48\textwidth}
         \centering
         \includegraphics[width=\textwidth]{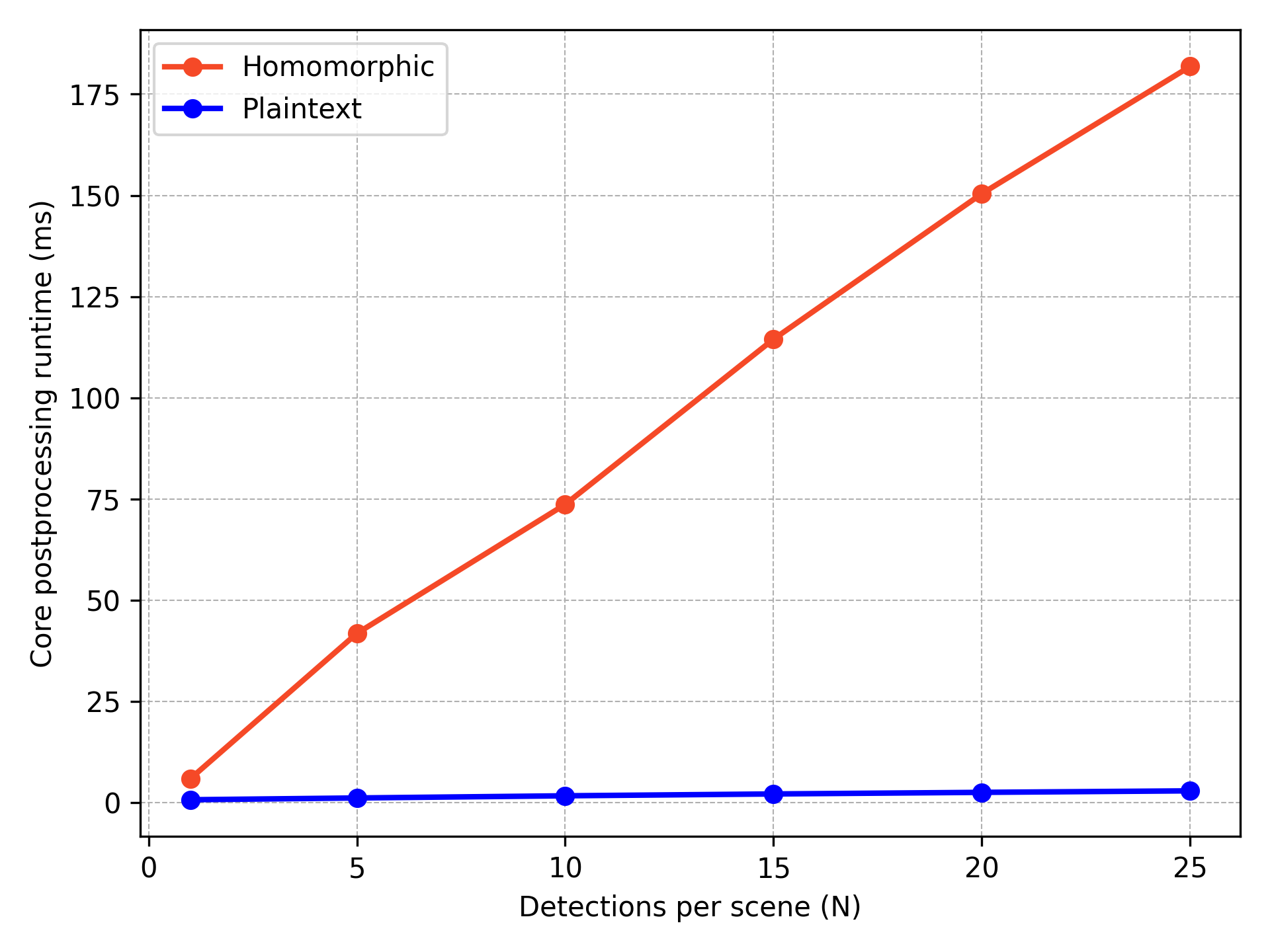}
         \caption{Comparison of core vendor postprocessing runtime between homomorphic and plaintext execution. Homomorphic processing exhibits linear scaling with a higher constant factor, while plaintext remains near-negligible.}
         \label{fig-exp-vendor-post-postprocessing}
     \end{subfigure}
        \caption{Vendor-side postprocessing performance. Homomorphic processing introduces significant computational overhead due to ciphertext operations, while preserving linear scaling with respect to the number of occupied bins.}
        \label{fig-exp-vendor-post-graphs}
\end{figure}
The vendor-side postprocessing results further validate the linear scaling behavior of the proposed pipeline. As shown in Figure~\ref{fig-exp-vendor-post-graphs}, the total homomorphic runtime grows approximately linearly with the number of detections, which directly corresponds to the number of occupied bins. The stage-wise breakdown reveals that ciphertext decryption dominates the computational cost, while moment-to-geometry transformation and bin merging contribute only marginal overhead.

In contrast, plaintext postprocessing remains negligible across all detection counts, highlighting that the additional cost is entirely attributable to homomorphic operations rather than the underlying fusion logic. Importantly, despite the large constant-factor overhead, the absence of super-linear growth confirms that the system scales predictably as $O(B)$, making it suitable for deployment under bounded scene complexity.

This result is particularly significant, as it demonstrates that the dominant cost arises from unavoidable cryptographic primitives rather than algorithmic inefficiencies, preserving scalability despite strong privacy guarantees.

\subsection{Situational Awareness Evaluation}
\label{sec:awareness}
In cooperative perception settings, the primary objective is to maximize the coverage of objects in the scene by aggregating complementary detections across heterogeneous perception systems. Hence, we evaluate the \textit{situational awareness}, rather than conventional detection accuracy. We conduct experiments on the KITTI~\cite{kitti} object detection benchmark and focus on the \textit{Car} and \textit{Pedestrian} classes, we use training split with corresponding annotations. The dataset provides synchronized RGB images and LiDAR point clouds, enabling evaluation of multi-modal perception and fusion. Camera-based detections are obtained using YOLOv8~\cite{redmon2016you, yolo-ultralytics}, while LiDAR-based detections are produced using PointPillars~\cite{pointpillars} and PV-RCNN~\cite{pvrcnn}. Evaluation is done on three configurations: (i) LiDAR-only baseline, (ii) \textsf{Sarus}, and (iii) an \textit{Upper Bound} corresponding to the union of camera and LiDAR detections. The LiDAR models are implemented using the OpenPCDet framework~\cite{openpcdet2020}, which provides standardized training and inference pipelines for 3D object detection on KITTI. Pretrained models are used to ensure consistent and reproducible performance. We define \textit{coverage} as the fraction of ground-truth objects that are matched by at least one detection. Matching is determined using an Intersection-over-Union (IoU) threshold. We report results at IoU at $0.3$, which provides a more appropriate measure of situational awareness while still requiring meaningful spatial overlap. For completeness, we also observe that higher IoU thresholds lead to lower apparent coverage due to reconstruction effects, rather than failure to detect objects.

To analyze the spatial behavior of fusion, we stratify evaluation by object distance into three non-overlapping bins: $[0,20)$ m, $[20,40)$ m, and $\geq 40$ m. This partitioning captures the distance dependent sensing characteristics of camera and LiDAR modalities. Specifically, LiDAR performance degrades with increasing range due to reduced point cloud density and sparsity, whereas camera-based detection remains comparatively robust. This allows us to isolate regimes of modality complementarity and quantify the effectiveness of fusion under varying sensing conditions.
\begin{figure}[t]
     \centering
     \begin{subfigure}[b]{0.49\textwidth}
         \centering
         \includegraphics[width=\textwidth]{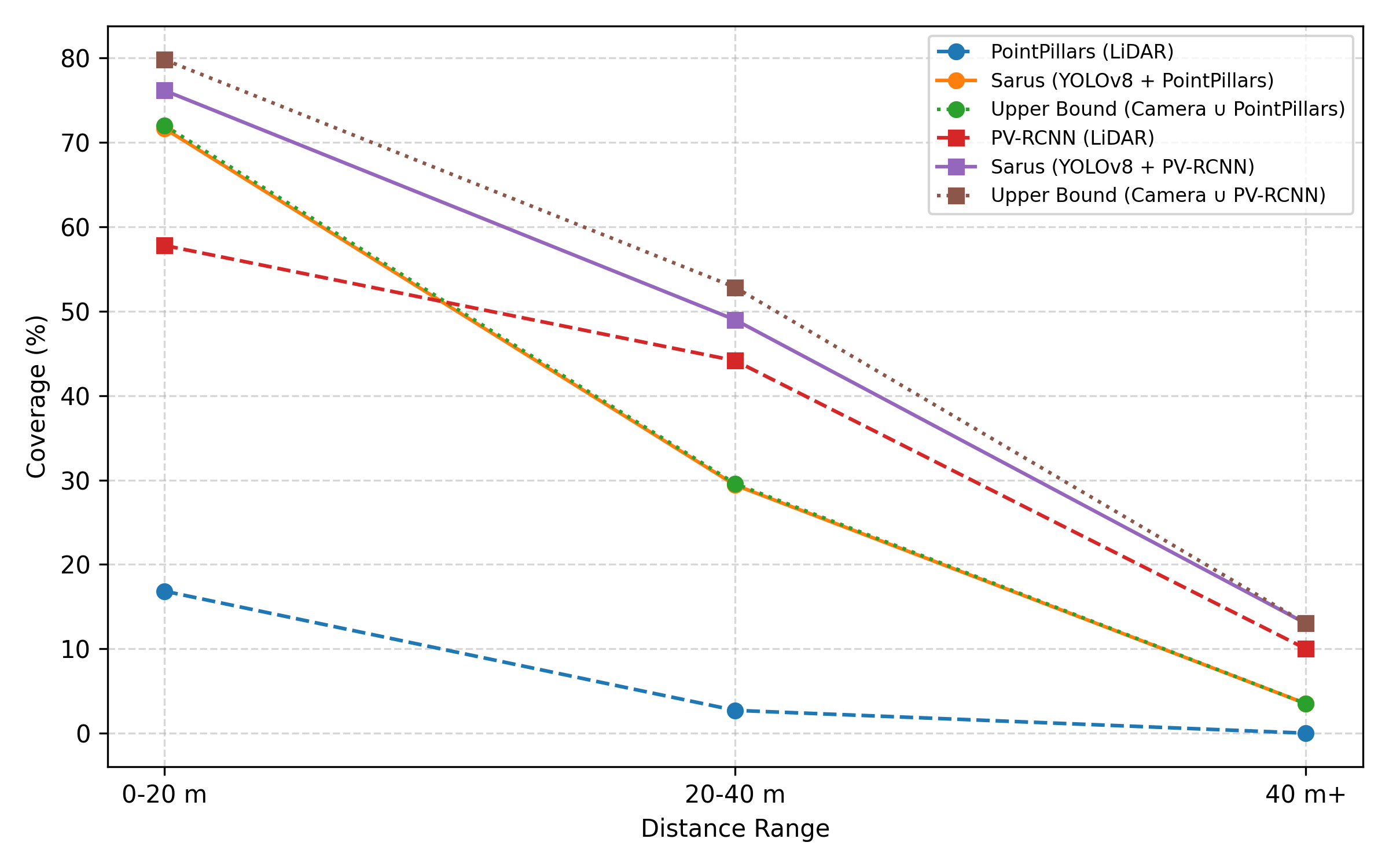}
         \caption{\textbf{Pedestrian:} Significant improvement in coverage when paired with PointPillars, and preserves performance with PV-RCNN, with gains primarily in near and mid ranges.}
         \label{fig-exp-kitti-pedestrian}
     \end{subfigure}
     \hfill
     \begin{subfigure}[b]{0.49\textwidth}
         \centering
         \includegraphics[width=\textwidth]{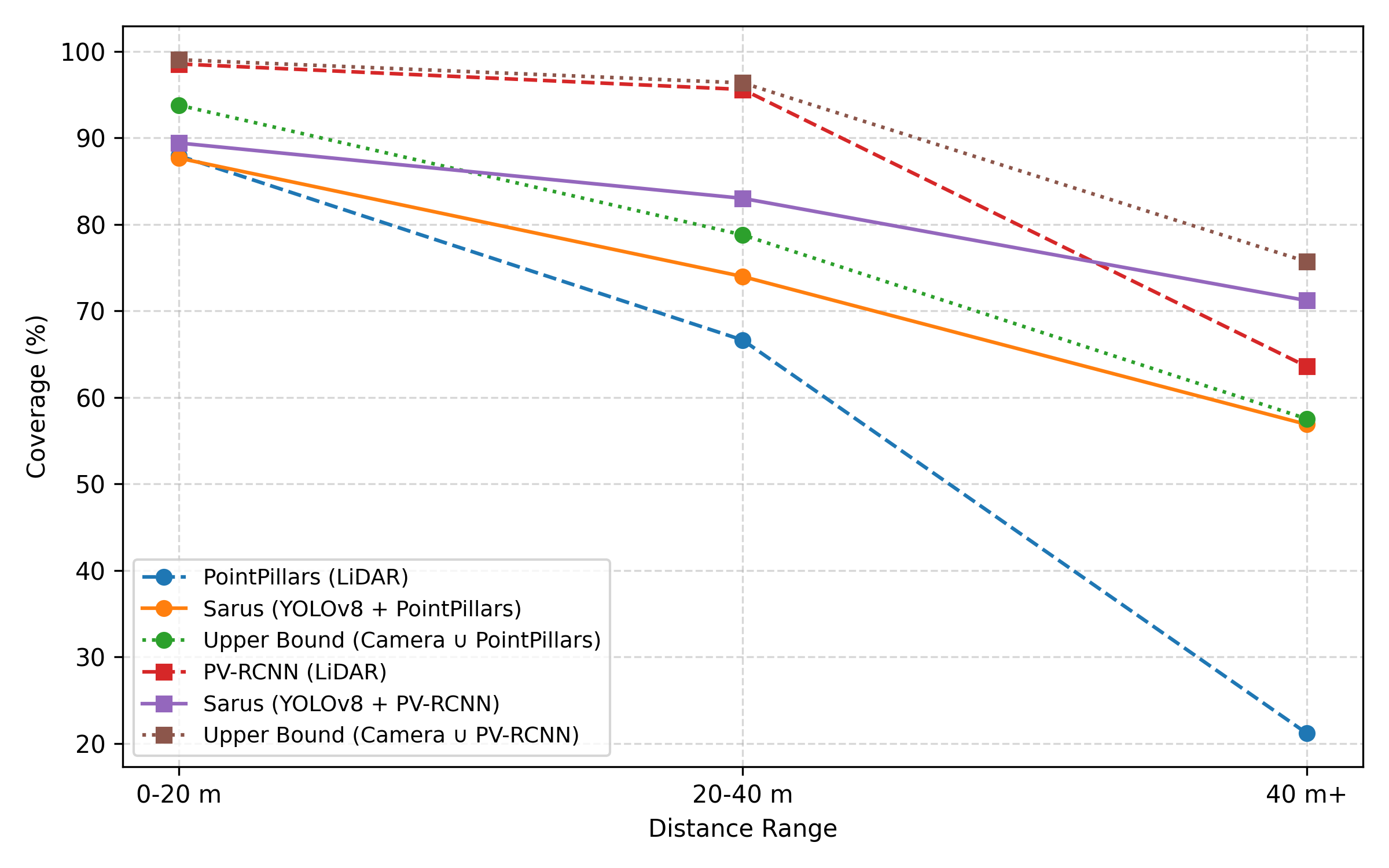}
         \caption{\textbf{Car:} Substantial improvement for PointPillars and yields complementary gains at longer ranges even with PV-RCNN, reflecting distance-dependent modality complementarity.}
         \label{fig-exp-kitti-car}
     \end{subfigure}
        \caption{Distance-based coverage analysis (IoU = $0.3$). Coverage is reported across three distance ranges: $[0, 20)$ m, $[20, 40)$ m, and $\geq 40$ m. \textsf{Sarus} improves situational awareness by aggregating complementary detections from camera and LiDAR modalities. Gains are most pronounced when  detectors operate in challenging regimes, such as long-range perception and LiDAR performance degrades due to sparsity.}
        \label{fig-exp-kitti}
\end{figure}

The results in Figure~\ref{fig-exp-kitti} illustrate the distance dependent behavior of the fusion framework across both Pedestrian and Car classes. When paired with PointPillars (see Figure~\ref{fig-exp-kitti-pedestrian} and Figure~\ref{fig-exp-kitti-car}), the framework leads to substantial improvements in coverage across all distance ranges, particularly for pedestrians and at longer distances where LiDAR-based detection degrades due to reduced point cloud density. In contrast, when combined with PV-RCNN, which achieves higher baseline coverage, \textsf{Sarus} largely preserves the existing performance while recovering a portion of the remaining missed detections. Notably, even in this setting, consistent gains are observed at longer ranges for cars, highlighting the role of cross-modal complementarity under challenging sensing conditions. These results indicate that the effectiveness of fusion is governed by both the baseline detector performance and the distance-dependent sensing characteristics of the modalities. Overall, framework effectively aggregates complementary information across camera and LiDAR inputs, improving situational awareness while maintaining a compact, privacy-preserving representation.

% \section{Discussion and Future Work}
% Framework relies on Gaussian moment aggregation over spatial bins to enable efficient and privacy-preserving fusion. While this representation preserves object-level awareness, it introduces approximation during reconstruction, which can degrade precise bounding box localization. This is reflected in reduced performance at higher IoU thresholds. Future work will explore adaptive binning strategies and higher-order representations to better preserve geometric fidelity without sacrificing efficiency. \textsf{Sarus} assumes that participating vendors provide honest detection outputs. However, in multi-vendor or adversarial settings, this assumption may not hold. An important extension is to integrate with cryptographic commitment and verification mechanisms~\cite{hasan2026hermessealzeroknowledgeassurance} to enable vendors to produce verifiable proofs of correct inference without revealing proprietary model details. This would ensure that fused outputs are derived from authentic detections, strengthening trust in cooperative perception.

\section{Limitations and Future Work}
\label{sec:limitations}

\paragraph{Approximate Representation and Localization:} \textsf{Sarus} relies on Gaussian moment aggregation over spatial bins to enable efficient and privacy-preserving fusion. This representation requires selecting class-specific spatial anchors $S$ and strides $s$, defined with respect to a common reference frame $\mathcal{F}_{\text{common}}$ (see Definition~\ref{def:common_frame}). These parameters determine how detections are assigned to bins and therefore influence both communication cost and reconstruction fidelity. While the moment-based representation preserves object-level awareness, it introduces approximation during binning and reconstruction, which can degrade precise bounding-box localization. This effect is reflected in reduced performance at higher IoU thresholds. Future work will explore adaptive choices of $S$ and $s$, learned or data-dependent binning strategies, and higher-order representations to better preserve geometric fidelity without sacrificing efficiency.

\paragraph{Dependence on Detection Quality:} The effectiveness of \textsf{Sarus} is influenced by the quality and complementarity of the underlying perception models. When baseline detectors already achieve high coverage, the relative gains from fusion may diminish. Conversely, larger improvements are expected when modalities or vendors exhibit complementary failure modes. A promising direction is to incorporate confidence-aware or reliability-weighted fusion mechanisms that adapt dynamically to detector performance.

\paragraph{Admission, Compliance, and Verifiable Inputs:} \textsf{Sarus} currently assumes that participating vendors submit payloads that are admissible for fusion: they follow the agreed schema, originate from legitimate participants, and conform to the public perception-fusion specification. In multi-vendor or adversarial settings, this assumption may not hold automatically. An important extension is to integrate Sarus with an admission and compliance layer based on credentialed vehicular communication, proof-carrying data, or succinct zero-knowledge compliance proofs. For example, a Hermes Seal--style mechanism could allow vendors to attach evidence that their submitted encrypted payloads satisfy a public specification without revealing raw sensor data, private detections, or proprietary model details. This would strengthen trust in the cooperative perception pipeline while preserving the modular role of Sarus as the encrypted fusion layer.

\paragraph{Key Management:} In our current formulation, the homomorphic encryption context, public encryption material, and decryption authority are assumed to be established before fusion begins. This leaves open important system-design questions, including which party generates the homomorphic encryption keys, which participants are authorized to decrypt the fused result, how secret keys are protected, and how key rotation or revocation should be handled when vendors join or leave the cooperative perception group. Future work will investigate threshold and multi-key variants of homomorphic encryption, where decryption of the fused output requires participation from multiple authorized parties rather than a single trusted key holder. Another promising direction is to combine encrypted aggregation with secure multi-party computation (MPC), enabling stronger control over key custody, joint decryption, and access policies for fused perception outputs.

\paragraph{Limited Modalities and Datasets:} Our evaluation focuses on camera and LiDAR fusion on the KITTI dataset. While this provides a controlled and reproducible benchmark, real-world deployments may involve additional modalities, such as radar, and more diverse operating environments. Extending Sarus to broader multi-modal and multi-dataset settings, including large-scale autonomous driving benchmarks, is an important direction for future work.

\paragraph{System-Level Integration:} \textsf{Sarus} is currently evaluated as a perception-level fusion module. Integrating it into full autonomous driving stacks, including tracking, prediction, planning, and decision-making, remains an open challenge. Future work will investigate how improved situational awareness from privacy-preserving fusion translates into downstream safety and robustness gains in closed-loop systems.

\section{Conclusion}
\label{sec:conclusion}
In this work, we presented \textsf{Sarus}, a privacy-preserving framework for multi-vendor cooperative perception that enables secure aggregation of inference-time detection outputs using CKKS-based homomorphic encryption. By introducing a moment-based representation over a shared spatial lattice, \textsf{Sarus} enables efficient fusion of structured perception outputs without requiring access to raw detections or proprietary model information. The proposed design eliminates the need for plaintext sharing, addressing key privacy and confidentiality challenges in multi-vendor perception systems.

We showed that the framework achieves scalable performance through spatial binning, with vendor payload construction scaling linearly with the number of detections and server-side fusion scaling as $O(BV)$ with respect to the number of occupied bins and vendors. Experimental results demonstrate that homomorphic encryption introduces only a bounded constant-factor overhead, while maintaining linear scaling in practice. These findings indicate that privacy-preserving multi-vendor perception fusion is feasible for real-time deployment when statistical compression and spatial sparsity are jointly exploited.

More broadly, this work highlights the importance of inference-time privacy in collaborative AI systems, where sensitive outputs rather than training data must be protected. Future work includes extending the framework to stronger adversarial models, exploring alternative cryptographic primitives with improved efficiency, and integrating the approach into real-world autonomous driving and V2X systems.

%Bibliography
\bibliographystyle{unsrt}  
\bibliography{references} 

@inproceedings{cheon2017homomorphic,
  title={Homomorphic encryption for arithmetic of approximate numbers},
  author={Cheon, Jung Hee and Kim, Andrey and Kim, Miran and Song, Yongsoo},
  booktitle={International conference on the theory and application of cryptology and information security},
  pages={409--437},
  year={2017},
  organization={Springer}
}

@INPROCEEDINGS{chen2019cooperatively,
  author={Chen, Qi and Tang, Sihai and Yang, Qing and Fu, Song},
  booktitle={2019 IEEE 39th International Conference on Distributed Computing Systems (ICDCS)}, 
  title={{Cooper: Cooperative Perception for Connected Autonomous Vehicles Based on 3D Point Clouds}}, 
  year={2019},
  volume={},
  number={},
  pages={514-524},
  note={\url{https://doi.org/doi:10.1109/ICDCS.2019.00058}}
}

@INPROCEEDINGS{xu2021opencood,
  author={Xu, Runsheng and Xiang, Hao and Xia, Xin and Han, Xu and Li, Jinlong and Ma, Jiaqi},
  booktitle={2022 International Conference on Robotics and Automation (ICRA)}, 
  title={{OPV2V: An Open Benchmark Dataset and Fusion Pipeline for Perception with Vehicle-to-Vehicle Communication}}, 
  year={2022},
  volume={},
  number={},
  pages={2583-2589},
  note={\url{https://doi.org/doi: 10.1109/ICRA46639.2022.9812038}}
}

@ARTICLE{xiang2023multi,
  author={Xiang, Chao and Feng, Chen and Xie, Xiaopo and Shi, Botian and Lu, Hao and Lv, Yisheng and Yang, Mingchuan and Niu, Zhendong},
  journal={IEEE Intelligent Transportation Systems Magazine}, 
  title={{Multi-Sensor Fusion and Cooperative Perception for Autonomous Driving: A Review}}, 
  year={2023},
  volume={15},
  number={5},
  pages={36-58},
  note={\url{https://doi.org/doi:10.1109/MITS.2023.3283864}}
}

@ARTICLE{kim2014multivehicle,
  author={Kim, Seong-Woo and Qin, Baoxing and Chong, Zhuang Jie and Shen, Xiaotong and Liu, Wei and Ang, Marcelo H. and Frazzoli, Emilio and Rus, Daniela},
  journal={IEEE Transactions on Intelligent Transportation Systems}, 
  title={Multivehicle Cooperative Driving Using Cooperative Perception: Design and Experimental Validation}, 
  year={2015},
  volume={16},
  number={2},
  pages={663-680},
  note={\url{https://doi.org/doi:10.1109/TITS.2014.2337316}}
}

@misc{umtri_sip,
  title        = {Smart Intersection Project},
  author       = {{University of Michigan Transportation Research Institute}},
  year         = {2023},
  howpublished = {\url{https://sip.umtri.umich.edu/}},
  note         = {Accessed: 2026}
}

@article{chen2020improved,
  title={{Improved Techniques for Model Inversion Attacks }},
  author={Chen, Si and Jia, Ruoxi and Qi, Guo-Jun},
  year={2020}
}

@inproceedings{wu2016methodology,
  title={{A Methodology for Formalizing Model-Inversion Attacks}},
  author={Wu, Xi and Fredrikson, Matthew and Jha, Somesh and Naughton, Jeffrey F},
  booktitle={2016 IEEE 29th computer security foundations symposium (CSF)},
  pages={355--370},
  year={2016},
  organization={IEEE},
  note={\url{https://doi.org/doi:10.1109/CSF.2016.32}}
}

@misc{wiz,
  title        = {{Wiz Discovers Flaws in GenAI Models Enabling Customer Data Theft}},
  author       = {{Wiz}},
  year         = {2024},
  howpublished = {\url{https://www.infosecurity-magazine.com/news/wiz-discovers-flaws-generative-ai/}},
  note         = {Accessed: 2026}
}

@inproceedings{dibbo2023sok,
  title={{Sok: Model inversion attack landscape: Taxonomy, challenges, and future roadmap}},
  author={Dibbo, Sayanton V},
  booktitle={2023 IEEE 36th Computer Security Foundations Symposium (CSF)},
  pages={439--456},
  year={2023},
  organization={IEEE},
  note={\url{https://doi.org/doi:10.1109/CSF57540.2023.00027}}
}

@INPROCEEDINGS{boehme2020talkycars,
  author={Boehme, Martin and Stang, Marco and Muetsch, Ferdin and Sax, Eric},
  booktitle={2020 IEEE Intelligent Vehicles Symposium (IV)}, 
  title={{TalkyCars: A Distributed Software Platform for Cooperative Perception}}, 
  year={2020},
  volume={},
  number={},
  pages={701-707},
  note={\url{https://doi.org/doi:10.1109/IV47402.2020.9304630}}
}

@ARTICLE{dai2020hybrid,
  author={Dai, Bin and Xu, Fanglin and Cao, Yuanyuan and Xu, Yang},
  journal={IEEE Systems Journal}, 
  title={Hybrid Sensing Data Fusion of Cooperative Perception for Autonomous Driving With Augmented Vehicular Reality}, 
  year={2021},
  volume={15},
  number={1},
  pages={1413-1422},
  note={\url{https://doi.org/doi:10.1109/JSYST.2020.3007202}}
}

@inproceedings{wang2020v2vnet,
  title={{V2VNet: Vehicle-to-vehicle Communication for Joint Perception and Prediction}},
  author={Wang, Tsun-Hsuan and Manivasagam, Sivabalan and Liang, Ming and Yang, Bin and Zeng, Wenyuan and Urtasun, Raquel},
  booktitle={European conference on computer vision},
  pages={605--621},
  year={2020},
  organization={Springer},
  note={\url{
https://doi.org/10.48550/arXiv.2008.07519}}
}

@InProceedings{xu2022v2x,
    author="Xu, Runsheng
    and Xiang, Hao
    and Tu, Zhengzhong
    and Xia, Xin
    and Yang, Ming-Hsuan
    and Ma, Jiaqi",
    editor="Avidan, Shai
    and Brostow, Gabriel
    and Ciss{\'e}, Moustapha
    and Farinella, Giovanni Maria
    and Hassner, Tal",
    title={{V2X-ViT: Vehicle-to-Everything Cooperative Perception with Vision Transformer}},
    booktitle="Computer Vision -- ECCV 2022",
    year="2022",
    publisher="Springer Nature Switzerland",
    address="Cham",
    pages="107--124",
    note={\url{https://doi.org/10.48550/arXiv.2203.10638}}
}

@inproceedings{ma2024macp,
  title={{MACP: Efficient model adaptation for cooperative perception}},
  author={Ma, Yunsheng and Lu, Juanwu and Cui, Can and Zhao, Sicheng and Cao, Xu and Ye, Wenqian and Wang, Ziran},
  booktitle={Proceedings of the IEEE/CVF Winter Conference on Applications of Computer Vision},
  pages={3373--3382},
  year={2024},
  note={\url{
https://doi.org/10.48550/arXiv.2310.16870}}
}

@INPROCEEDINGS{rauch2012car2x,
  author={Rauch, Andreas and Klanner, Felix and Rasshofer, Ralph and Dietmayer, Klaus},
  booktitle={2012 IEEE Intelligent Vehicles Symposium}, 
  title={{Car2X-based perception in a high-level fusion architecture for cooperative perception systems}}, 
  year={2012},
  volume={},
  number={},
  pages={270-275},
  note={\url{
https://doi.org/10.1109/IVS.2012.6232130}}
}

@inproceedings{zheng2022robust,
  title={A robust strategy for roadside cooperative perception based on multi-sensor fusion},
  author={Zheng, Shaowu and Xie, Chong and Yu, Shanhu and Ye, Ming and Huang, Ruyi and Li, Weihua},
  booktitle={2022 International Conference on Sensing, Measurement \& Data Analytics in the era of Artificial Intelligence (ICSMD)},
  pages={1--6},
  year={2022},
  organization={IEEE},
  note={\url{https://doi.org/10.1109/ICSMD57530.2022.10058282}}
}

@ARTICLE{arnold2020cooperative,
  author={Arnold, Eduardo and Dianati, Mehrdad and de Temple, Robert and Fallah, Saber},
  journal={IEEE Transactions on Intelligent Transportation Systems}, 
  title={Cooperative Perception for 3D Object Detection in Driving Scenarios Using Infrastructure Sensors}, 
  year={2022},
  volume={23},
  number={3},
  pages={1852-1864},
 note={\url{https://doi.org/10.1109/TITS.2020.3028424}}
}

@ARTICLE{wei2025cooperative,
  author={Wei, Chuheng and Wu, Guoyuan and Barth, Matthew J.},
  journal={Proceedings of the IEEE}, 
  title={{Cooperative Perception for Automated Driving: A Survey of Algorithms, Applications, and Future Directions}}, 
  year={2025},
  volume={},
  number={},
  pages={1-27},
  note={\url{https://doi.org/10.1109/JPROC.2025.3608874}}
}

@ARTICLE{zhang2024collaborative,
  author={Zhang, Lei and Wang, Binglu and Zhao, Yongqiang and Yuan, Yuan and Zhou, Tianfei and Li, Zhijun},
  journal={IEEE Transactions on Cybernetics}, 
  title={{Collaborative Multimodal Fusion Network for Multiagent Perception}}, 
  year={2025},
  volume={55},
  number={1},
  pages={486-498},
  note={\url{https://doi.org/10.1109/TCYB.2024.3491756}}
}

@ARTICLE{zhou2024vit,
  author={Zhou, Yang and Yang, Cai and Wang, Ping and Wang, Chao and Wang, Xinhong and Ngoc Van, Nguyen},
  journal={IEEE Access}, 
  title={{ViT-FuseNet: Multimodal Fusion of Vision Transformer for Vehicle-Infrastructure Cooperative Perception}}, 
  year={2024},
  volume={12},
  number={},
  pages={31640-31651},
  note={\url{https://doi.org/10.1109/ACCESS.2024.3368404}}
}

@ARTICLE{he2025mdnet,
  author={He, Junyang and Deng, Xiaoheng and Gui, Jinsong and Zhang, Tao and He, Xiangjian},
  journal={IEEE Internet of Things Journal}, 
  title={{MDNet: Multimodal Cooperative Perception via Spatial Alignment of Modal Decision-Making}}, 
  year={2025},
  volume={12},
  number={11},
  pages={16142-16154},
   note={\url{https://doi.org/10.1109/JIOT.2025.3531145}}
}

@ARTICLE{yin2023v2vformer++,
  author={Yin, Hongbo and Tian, Daxin and Lin, Chunmian and Duan, Xuting and Zhou, Jianshan and Zhao, Dezong and Cao, Dongpu},
  journal={IEEE Transactions on Intelligent Transportation Systems}, 
  title={{V2VFormer++: Multi-Modal Vehicle-to-Vehicle Cooperative Perception via Global-Local Transformer}}, 
  year={2024},
  volume={25},
  number={2},
  pages={2153-2166},
  note={\url{https://doi.org/10.1109/TITS.2023.3314919}}
}

@INPROCEEDINGS{zhang2022multi,
  author={Zhang, Hui and Luo, Guiyang and Cao, Yuanzhouhan and Jin, Yi and Li, Yidong},
  booktitle={2022 IEEE 13th International Symposium on Parallel Architectures, Algorithms and Programming (PAAP)}, 
  title={{Multi-Modal Virtual-Real Fusion based Transformer for Collaborative Perception}}, 
  year={2022},
  volume={},
  number={},
  pages={1-6},
  note={\url{https://doi.org/10.1109/PAAP56126.2022.10010640}}
}

@inproceedings{li2025rg,
  title={{RG-Attn: Radian Glue Attention for Multi-modal Multi-agent Cooperative Perception}},
  author={Li, Lantao and Yang, Kang and Zhang, Wenqi and Wang, Xiaoxue and Sun, Chen},
  booktitle={Proceedings of the IEEE/CVF International Conference on Computer Vision},
  pages={1763--1772},
  year={2025},
  note={\url{
https://doi.org/10.48550/arXiv.2501.1680}}
}

@ARTICLE{li2023learning,
  author={Li, Jinlong and Xu, Runsheng and Liu, Xinyu and Ma, Jin and Chi, Zicheng and Ma, Jiaqi and Yu, Hongkai},
  journal={IEEE Transactions on Intelligent Vehicles}, 
  title={{Learning for Vehicle-to-Vehicle Cooperative Perception Under Lossy Communication}}, 
  year={2023},
  volume={8},
  number={4},
  pages={2650-2660},
  note={\url{https://doi.org/10.1109/TIV.2023.3260040}}
}

@INPROCEEDINGS{jiang2025multimodal,
  author={Jiang, Hanwen and Zhou, Shijun and Zhu, Konglin and Andrzejak, Artur and Gong, Yi},
  booktitle={2025 9th IEEE International Conference on Network Intelligence and Digital Content (IC-NIDC)}, 
  title={{A Multimodal Collaborative Perception Framework in Challenging Environments}}, 
  year={2025},
  volume={},
  number={},
  pages={62-66},
  note={\url{https://doi.org/10.1109/IC-NIDC67200.2025.11390536}}
}

@article{mammen2021federated,
  title={{Federated Learning: Opportunities and Challenges}},
  author={Mammen, Priyanka Mary},
  journal={arXiv preprint arXiv:2101.05428},
  year={2021},
  note={\url{
https://doi.org/10.48550/arXiv.2101.05428}}
}

@inproceedings{mcmahan2017communication,
  title={{Communication-Efficient Learning of Deep Networks from Decentralized Data}},
  author={McMahan, Brendan and Moore, Eider and Ramage, Daniel and Hampson, Seth and y Arcas, Blaise Aguera},
  booktitle={Artificial intelligence and statistics},
  pages={1273--1282},
  year={2017},
  organization={Pmlr},
  note={\url{https://doi.org/10.48550/arXiv.1602.05629}}
}

@article{kairouz2021advances,
  title={{Advances and Open Problems in Federated Learning}},
  author={Kairouz, Peter and McMahan, H Brendan},
  journal={Foundations and trends in machine learning},
  volume={14},
  number={1-2},
  pages={1--210},
  year={2021},
  publisher={Emerald Publishing Limited},
  note={\url{
https://doi.org/10.48550/arXiv.1912.04977}}
}

@inproceedings{karimireddy2020scaffold,
  title={{SCAFFOLD: Stochastic Controlled Averaging for Federated Learning}},
  author={Karimireddy, Sai Praneeth and Kale, Satyen and Mohri, Mehryar and Reddi, Sashank and Stich, Sebastian and Suresh, Ananda Theertha},
  booktitle={International conference on machine learning},
  pages={5132--5143},
  year={2020},
  organization={PMLR},
  note={\url{
https://doi.org/10.48550/arXiv.1910.06378
}}
}

@ARTICLE{zhang2024federated,
  author={Zhang, Zhenrong and Liu, Jianan and Zhou, Xi and Huang, Tao and Han, Qing-Long and Liu, Jingxin and Liu, Hongbin},
  journal={IEEE Robotics and Automation Letters}, 
  title={{On the Federated Learning Framework for Cooperative Perception}}, 
  year={2024},
  volume={9},
  number={11},
  pages={9423-9430},
  note={\url{https://doi.org/10.1109/LRA.2024.3457374}}
}

@ARTICLE{abdel2021vehicular,
  author={Abdel-Aziz, Mohamed K. and Perfecto, Cristina and Samarakoon, Sumudu and Bennis, Mehdi and Saad, Walid},
  journal={IEEE Transactions on Communications}, 
  title={{Vehicular Cooperative Perception Through Action Branching and Federated Reinforcement Learning}}, 
  year={2022},
  volume={70},
  number={2},
  pages={891-903},
  note={\url{https://doi.org/10.1109/TCOMM.2021.3126650}}
}

@inproceedings{lu2025privacy,
  title={{Privacy-Preserving V2X Collaborative Perception Integrating Unknown Collaborators}},
  author={Lu, Bin and Xiao, Xinyu and Zhang, Changzhou and Zhou, Yang and Xiang, Zhiyu and Shan, Hangguan and Liu, Eryun},
  booktitle={Proceedings of the AAAI Conference on Artificial Intelligence},
  volume={39},
  number={6},
  pages={5802--5810},
  year={2025}
}

@article{hesamifard2017cryptodl,
  title={{CryptoDL: Deep Neural Networks over Encrypted Data}},
  author={Hesamifard, Ehsan and Takabi, Hassan and Ghasemi, Mehdi},
  journal={arXiv preprint arXiv:1711.05189},
  year={2017},
  note={\url{https://doi.org/10.48550/arXiv.1711.05189}}
}

@INPROCEEDINGS{xu2019cryptonn,
  author={Xu, Runhua and Joshi, James B.D. and Li, Chao},
  booktitle={2019 IEEE 39th International Conference on Distributed Computing Systems (ICDCS)}, 
  title={CryptoNN: Training Neural Networks over Encrypted Data}, 
  year={2019},
  volume={},
  number={},
  pages={1199-1209},
  note={\url{https://doi.org/10.1109/ICDCS.2019.00121}}
}

@misc{cryptoeprint:2014/331,
      author = {Raphael Bost and Raluca Ada Popa and Stephen Tu and Shafi Goldwasser},
      title = {{Machine Learning Classification over Encrypted Data}},
      howpublished = {Cryptology {ePrint} Archive, Paper 2014/331},
      year = {2014},
      note={\url{https://doi.org/10.14722/ndss.2015.23241}} 
}

@inproceedings{moon2025thor,
  title={{THOR: Secure Transformer Inference with Homomorphic Encryption}},
  author={Moon, Jungho and Yoo, Dongwoo and Jiang, Xiaoqian and Kim, Miran},
  booktitle={Proceedings of the 2025 ACM SIGSAC Conference on Computer and Communications Security},
  pages={3765--3779},
  year={2025},
  note={\url{https://doi.org/10.1145/3719027.3765150}}
}

@ARTICLE{kim2023optimized,
  author={Kim, Dongwoo and Guyot, Cyril},
  journal={IEEE Transactions on Information Forensics and Security}, 
  title={{Optimized Privacy-Preserving CNN Inference With Fully Homomorphic Encryption}}, 
  year={2023},
  volume={18},
  number={},
  pages={2175-2187},
  note={\url{https://doi.org/110.1109/TIFS.2023.3263631}}
}

@article{vassilev2026assessment,
  title={{On the Assessment of Sensitivity of Autonomous Vehicle Perception}},
  author={Vassilev, Apostol and Hasan, Munawar and Griffor, Edward and Jin, Honglan and Piliptchak, Pavel and Arora, Mahima and Gamage, Thoshitha},
  journal={arXiv preprint arXiv:2602.00314},
  year={2026},
  note={\url{https://doi.org/10.48550/arXiv.2602.00314}}
}

@online{vtti,
  title = {{Virginia Tech Transportation Institute}},
  author = {VTTI},
  url = {https://www.vtti.vt.edu/},
  urldate = {2026-03-23},
}

@misc{kitti,
  title        = {{KITTI Vision Benchmark Suite}},
  author       = {{KITTI}},
  year         = {2012},
  howpublished = {\url{https://www.cvlibs.net/datasets/kitti/}},
  note         = {Accessed: 2026}
}

@article{pointpillars,
  title={{PointPillars: Fast Encoders for Object Detection from Point Clouds}},
  author={Alex H. Lang and Sourabh Vora and Holger Caesar and Lubing Zhou and Jiong Yang and Oscar Beijbom},
  year={2019},
  note={\url{
https://doi.org/10.48550/arXiv.1812.05784}}
}

@article{pvrcnn,
  title={{PV-RCNN: Point-Voxel Feature Set Abstraction for 3D Object Detection}},
  author={Shaoshuai Shi and Chaoxu Guo and Li Jiang and Zhe Wang and Jianping Shi and Xiaogang Wang and Hongsheng Li},
  year={2021},
  note={\url{
https://doi.org/10.48550/arXiv.1912.13192}}
}

@INPROCEEDINGS{redmon2016you,
  author={Redmon, Joseph and Divvala, Santosh and Girshick, Ross and Farhadi, Ali},
  booktitle={2016 IEEE Conference on Computer Vision and Pattern Recognition (CVPR)}, 
  title={{You Only Look Once: Unified, Real-Time Object Detection}}, 
  year={2016},
  volume={},
  number={},
  pages={779-788},
  note={\url{https://doi.org/10.1109/CVPR.2016.91}}}

@online{yolo-ultralytics,
  title = {{Ultralytics YOLO}},
  author = {Ultralytics},
  url = {https://docs.ultralytics.com/models/yolov8/},
   note         = {Accessed: 2026}
}

@misc{openpcdet2020,
    title={{OpenPCDet: An Open-source Toolbox for 3D Object Detection from Point Clouds}},
    author={{OpenPCDet Development Team}},
    howpublished = {\url{https://github.com/open-mmlab/OpenPCDet}},
    year={2020},
    note         = {Accessed: 2026}
}

@article{hasan2026hermessealzeroknowledgeassurance,
  title={{Hermes Seal: Zero-Knowledge Assurance for Autonomous Vehicle Communications}},
  author={Hasan, Munawar and Vassilev, Apostol and Griffor, Edward and Gamage, Thoshitha},
  journal={arXiv preprint arXiv:2603.26343},
  year={2026},
  note={\url{
https://doi.org/10.48550/arXiv.2603.26343}}
}

@misc{vassilev2025,
author={Vassilev, Apostol and Oprea, Alina  and Fordyce, Alice and Anderson, Hyrum and Davies,  Xander  and Hamin, Maia},
year={2025},
title={Adversarial Machine Learning: A
Taxonomy and Terminology of Attacks and Mitigations},
note={National Institute of Standards and Technology Gaithersburg, MD, NIST Trustworthy and Responsible AI, NIST AI 100-2e2025 https://doi.org/10.6028/NIST.AI.100-2e2025}
}

@inproceedings{chiesa2010pcd,
  title     = {Proof-Carrying Data and Hearsay Arguments from Signature Cards},
  author    = {Chiesa, Alessandro and Tromer, Eran},
  booktitle = {Innovations in Computer Science (ICS)},
  year      = {2010}
}

@standard{ieee1609,
  title        = {{IEEE Standard for Wireless Access in Vehicular Environments---Security Services for Application and Management Messages}},
  organization = {IEEE},
  number       = {IEEE Std 1609.2},
  year         = {2022}
}

@article{goldwasser1984probabilistic,
  title={Probabilistic Encryption},
  author={Goldwasser, Shafi and Micali, Silvio},
  journal={Journal of Computer and System Sciences},
  volume={28},
  number={2},
  pages={270--299},
  year={1984},
  publisher={Elsevier}
}

@book{katz2020introduction,
  title={Introduction to Modern Cryptography},
  author={Katz, Jonathan and Lindell, Yehuda},
  edition={3},
  year={2020},
  publisher={CRC Press}
}

@article{elfes1989occupancy,
  title={Using occupancy grids for mobile robot perception and navigation},
  author={Elfes, Alberto},
  journal={Computer},
  volume={22},
  number={6},
  pages={46--57},
  year={1989},
  publisher={IEEE}
}

@article{phan2018calibrating,
  title={Calibrating Uncertainties in Object Localization Task}}

@article{wang2020inferring,
  title={{Inferring Spatial Uncertainty in Object Detection}},
  author={Wang, Zining and Feng, Di and Zhou, Yiyang and Rosenbaum, Lars and Timm, Fabian and Dietmayer, Klaus and Tomizuka, Masayoshi and Zhan, Wei},
  journal={arXiv preprint arXiv:2003.03644},
  year={2020}
}

\end{document}